\documentclass[reprint,amsmath,amssymb,aps,prx]{revtex4-2}

\usepackage{comment}

\usepackage{graphicx}
\usepackage{dcolumn}
\usepackage{bm}
\usepackage{braket}
\usepackage{amsmath}
\usepackage{mathtools}
\usepackage{subfig}
\usepackage{caption}
\begin{document}


\title{Construction of Antisymmetric Variational Quantum States \\ with Real-Space Representation}

\author{Takahiro Horiba}
\email{t-horiba@mosk.tytlabs.co.jp}
\author{Soichi Shirai}
\author{Hirotoshi Hirai}
\affiliation{Toyota Central R\&D Labs., Inc., 41-1 Yokomichi, Nagakute, Aichi 480-1192, Japan}

\date{\today}

\begin{abstract}

Electronic state calculations using quantum computers are mostly based on second quantization, which is suitable for qubit representation.
Another way to describe electronic states on a quantum computer is first quantization, 
which is expected to achieve smaller scaling with respect to the number of basis functions than second quantization. 
Among basis functions, a real-space basis is an attractive option for quantum dynamics simulations in the fault-tolerant quantum computation (FTQC) era.
A major difficulty in first quantization with a real-space basis is state preparation for many-body electronic systems. 
This difficulty stems from of the antisymmetry of electrons, 
and it is not straightforward to construct antisymmetric quantum states on a quantum circuit.

In the present paper, we provide a design principle for constructing a variational quantum circuit to prepare an antisymmetric quantum state. 
The proposed circuit generates the superposition of exponentially many Slater determinants, that is, a multi-configuration state, 
which provides a systematic approach to approximating the exact ground state.
We implemented the variational quantum eigensolver (VQE) to obtain the ground state of a one-dimensional hydrogen molecular system.
As a result, the proposed circuit well reproduced the exact antisymmetric ground state and its energy, 
whereas the conventional variational circuit yielded neither an antisymmetric nor a symmetric state.
Furthermore, we analyzed the many-body wave functions based on quantum information theory, 
which illustrated the relation between the electron correlation and the quantum entanglement.

\end{abstract}

\maketitle

\section{Introduction}

Quantum computers are currently attracting increasing attention as promising hardware for materials computations~\cite{mcardle2020quantum, bauer2020quantum, ma2020quantum}, 
and a number of studies on a variety of material systems have been conducted
~\cite{nam2020ground, shirai2022calculation, shirai2023computational, hirai2023excited, colless2018computation, kassal2009quantum, rall2020quantum,kassal2008polynomial, oftelie2020towards, ollitrault2021molecular}.
Such materials computations are mostly based on second quantization~\cite{peruzzo2014variational, mcclean2016theory, cerezo2021variational}, 
which is suitable for describing electronic states on quantum computers.

An alternative way to describe electronic states on quantum computers is first quantization, 
in which a wave function is specified by the expansion coefficients of the basis functions.
With quantum computers, 
it is possible to obtain a wave function represented by an exponential number of basis functions with a polynomial number of qubits~\cite{zalka1998efficient}. 
Therefore, first quantization offers the possibility of achieving smaller scaling 
with respect to the number of basis functions than second quantization~\cite{abrams1997simulation, berry2018improved, babbush2019quantum, chan2023grid, su2021fault}.

Among basis functions, a real-space basis is an attractive option 
because a systematic improvement of computational accuracy and a rapid convergence to the continuum limit can be expected by increasing the number of qubits~\cite{hirose2005first}.
In addition, a real-space basis can be applied to systems with a variety of boundary conditions~\cite{ohba2012linear, hirose2005first}, 
and thus is suitable for quantum dynamics calculations~\cite{childs2022quantum}.
Recently, Chan et al.\ proposed quantum circuits for computing real-space quantum dynamics based on first quantization~\cite{chan2023grid}, 
which represented a promising blueprint of quantum simulations in the fault-tolerant quantum computing (FTQC) era.

However, first quantization has a significant challenge, namely state preparation.
Let us consider preparing the ground state of the system, 
which is a typical choice of initial state for dynamics calculations.
As a state preparation method, we consider quantum phase estimation (QPE)~\cite{nielsen2002quantum} 
which has been employed in several studies on first quantization~\cite{berry2018improved, chan2023grid}.
By using QPE, it is possible to distill the ground state from an input state which has sufficient overlap with the ground state~\cite{berry2018improved}.
Thus, the problem is how to prepare such an input state.

In first quantization with a real-space basis, 
preparing an input state takes a tremendous number of gate operations 
because probability the amplitudes of the many-body wave function need to be encoded into a state vector of a quantum circuit.
Although several amplitude-encoding methods have been proposed~\cite{holmes2020efficient, koppe2023amplitude, ollitrault2020nonadiabatic, nakaji2022approximate}, 
it is not at all straightforward to prepare an approximate ground state with any of them. 
This state preparation problem is often avoided by using oracle circuits.
Compared to constructing such oracle circuits, 
variational methods such as the variational quantum eigensolver (VQE)~\cite{peruzzo2014variational, mcclean2016theory, cerezo2021variational} 
are considered to be relatively feasible approaches. 
In fact, a number of studies have proposed preparing an input state using the VQE in second quantization~\cite{halder2023iterative,dong2022ground}.

Unfortunately, it is also not straightforward to implement the VQE in first quantization.
This is due to the antisymmetry of electrons, which are Fermi particles.
In second quantization, antisymmetry does not need to be considered for variational quantum states, 
because antisymmetry is naturally introduced as the anticommutation relations of creation and annihilation operators.
By contrast, in first quantization, antisymmetry is imposed on the many-body wave function itself, 
and thus variational quantum states must satisfy antisymmetry.
Nevertheless, there is no guarantee that quantum states generated by conventional variational quantum circuits will satisfy antisymmetry, 
which raises the possibility that the VQE will yield non-antisymmetric states.
Therefore, a variational quantum circuit that generates only antisymmetric quantum states is required in order to obtain the electronic ground state by the VQE.

In the present paper, we provide a design principle for constructing such an antisymmetrized variational quantum circuit.
Our proposed circuit consists of two types of variational circuits 
that achieve antisymmetry-preserving transformations of a state vector on a Hilbert space.
It is noteworthy that the proposed circuit generates a multi-configuration (MC) state. 
That is, it is possible to generate superpositions of an exponentially large number of Slater determinants 
by alternately layering these two types of variational circuits. 
This scheme provides a systematic approach to approximating the exact ground state.
To verify the validity of our method, we performed the VQE calculation for a one-dimensional hydrogen molecular (1D-$\rm{H_2}$) system, 
and demonstrated that the proposed circuit well reproduced the exact antisymmetric, or fermionic, ground state and its energy.
In addition to implementing the VQE, we analyzed the many-body wave functions based on quantum information theory.
Such an analysis reveals the microscopic electronic structure of a many-body wave function represented in real space 
and illustrates the relation between the electron correlation and the quantum entanglement.

\section{Method}

\begin{figure*}[!t]
  \centering
  \includegraphics[keepaspectratio, scale=0.45]{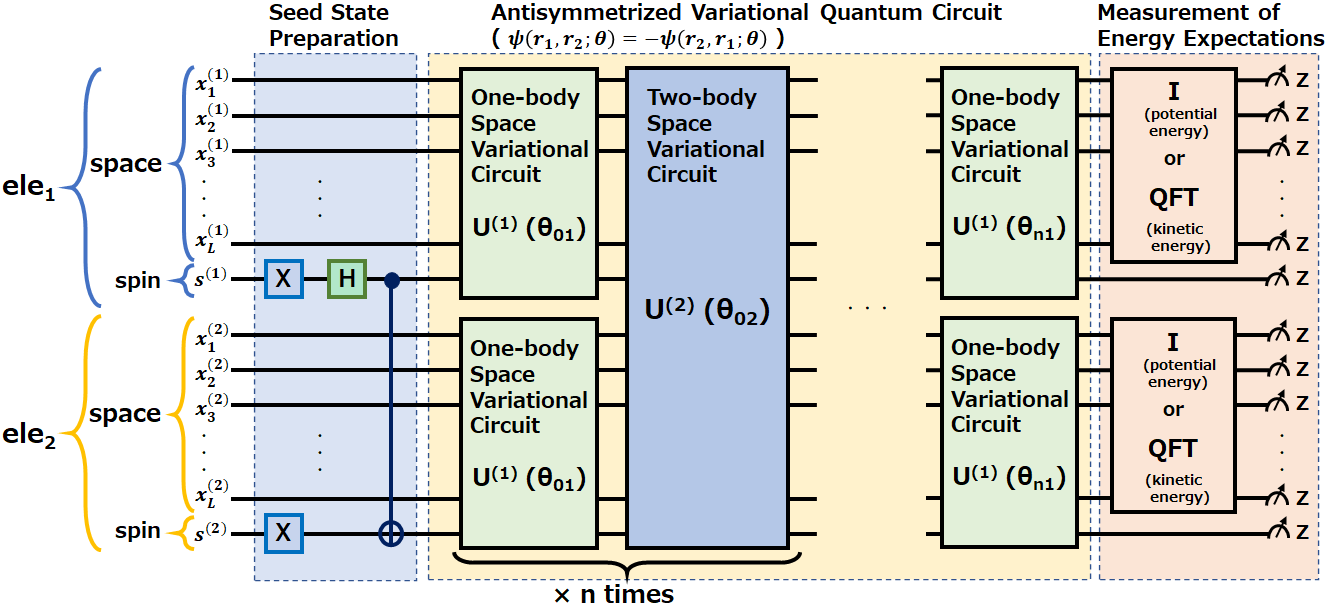}
  \caption{Quantum circuit to generate an antisymmetrized variational quantum state for one-dimensional two-electron systems}
  \label{fig:overall_circuit}
\end{figure*}

In this section, we introduce the first-quantized VQE framework based on the real-space representation 
and describe our proposed variational quantum circuit.
We also describe the setting of the numerical experiments for a 1D-$\rm{H_2}$ system.

\subsection{First-quantized VQE with a real-space basis}
To begin with, we briefly describe the first-quantized formulation of the many-body electron problem.
The many-body Schr\"{o}dinger equation for an $\eta$-electron molecular system is expressed by the following equation:
\begin{equation}
H\ket{\psi(\bm{r_1},\bm{r_2},\cdots,\bm{r_\eta})} = E\ket{\psi(\bm{r_1},\bm{r_2},\cdots,\bm{r_\eta})}, \label{eq:shrodinger}\\
\end{equation}
where $\ket{\psi(\bm{r_1},\bm{r_2},\cdots,\bm{r_\eta})}$ is the antisymmetric many-body wave function to be constructed on a quantum circuit.
The molecular Hamiltonian $H$ with the Born-Oppenheimer approximation is expressed in atomic units as follows:
\begin{equation}
H = \sum_{i=1}^{\eta}\left[-\frac{{\bm{\nabla_i}}^2}{2}-\sum_{p}\frac{Z_{p}}{|\bm{r_i}-\bm{R_p}|}\right]+\sum_{i<j}\frac{1}{|\bm{r_i}-\bm{r_j}|}, \label{eq:hamiltonian}\\  
\end{equation}
where $r_i$ is the $i$th electron coordinate, $R_p$ is the $p$th nuclear coordinate, and $Z_p$ is the atomic number of the $p$th nucleus.
The VQE in this study is based on this first-quantized Hamiltonian and the many-body wave function represented in real space.

Next, we introduce the real-space basis.
Let us consider the expansion of the many-body wave function by the real-space basis $\ket{\delta(\bm{r_1},\bm{r_2},\cdots,\bm{r_\eta})}$ (delta functions located at grid points $\bm{r_1},\bm{r_2},\cdots,\bm{r_\eta}$). 
The expansion coefficient of a many-body wave function is given as 
\begin{equation}
\psi(\bm{r'_1},\bm{r'_2},\cdots,\bm{r'_\eta}) = \braket{\delta(\bm{r'_1},\bm{r'_2},\cdots,\bm{r'_\eta})|\psi(\bm{r_1},\bm{r_2},\cdots,\bm{r_\eta})}.\\
\end{equation}
The real-space basis of an $\eta$-electron system consists of that of one-electron system $\ket{\delta(\bm{r_i})}$, as follows:
\begin{equation}
\ket{\delta(\bm{r_1},\bm{r_2},\cdots,\bm{r_\eta})} = \ket{\delta(\bm{r_1})}\ket{\delta(\bm{r_2})}\cdots\ket{\delta(\bm{r_\eta})}.\\
\end{equation}
$\ket{\delta(\bm{r_i})}$ consists of three-dimensional spatial coordinates $x, y, z$ and one spin coordinate $s$.
Assuming that $L$ qubits are assigned to each spatial dimension and one qubit to spin, $\ket{\delta(\bm{r_i})}$ is expressed as 
\begin{equation}
\ket{\delta(\bm{r_i})} = \ket{{{x_1}}^{(i)}\cdots{{x_L}}^{(i)}} \ket{{{y_1}}^{(i)}\cdots{{y_L}}^{(i)}} \ket{{{z_1}}^{(i)}\cdots{{z_L}}^{(i)}} \ket{s^{(i)}}, \\
\end{equation}
where ${x_k}^{(i)},{y_k}^{(i)},{z_k}^{(i)},s^{(i)} \in \left\{0,1\right\}, \forall k \in[1,L]$.
The number of qubits that constitute $\ket{\delta(\bm{r_1},\bm{r_2},\cdots,\bm{r_\eta})}$ is $\eta(3L+1)$, 
and thus the many-body wave function is expanded in $2^{\eta(3L+1)}$ basis functions.
Quantum computers are expected to realize such an exponentially large number of basis functions with a polynomial number of qubits, 
which is a significant advantage over classical computers.

In order to implement the VQE, it is necessary to measure the energy expectation value of a quantum state $\ket{\psi(\bm{r_1},\bm{r_2},\cdots,\bm{r_\eta})}$. 
The energy expectation value of molecular Hamiltonian $H$ in Eq.~\ref{eq:hamiltonian} is expressed as follows: 
\begin{equation}
\begin{split}
E &= \braket{\psi(\bm{r_1},\bm{r_2},\cdots,\bm{r_\eta})|H|\psi(\bm{r_1},\bm{r_2},\cdots,\bm{r_\eta})} \\
&= E_K + E_V^{e-n} + E_V^{e-e}, 
\end{split}
\end{equation}
where $E_K$ is the electron kinetic energy, $E_V^{e-n}$ is the electron-nuclei Coulomb attraction energy, and $E_V^{e-e}$ is the electron-electron Coulomb repulsion energy.
The kinetic and Coulomb energy operators of the Hamiltonian are diagonal in momentum space and real space, respectively.
The momentum space basis is obtained by the quantum Fourier transformation (QFT) of the real-space basis.
Letting $\ket{\psi(\bm{k_i})}=U_{\mathrm{QFT}}\ket{\psi(\bm{r_i})}$ be the one-body momentum space basis, the kinetic energy $E_K$ is expressed as 
\begin{equation}
E_K =\sum_{i=1}^{\eta} -\frac{{\bm{k_i}}^2}{2}|\psi(\bm{k_i})|^2. 
\end{equation}
The Coulomb energies $E_V^{e-n}, E_V^{e-e}$ are expressed in terms of one-body and two-body real-space bases as follows: 
\begin{equation}
\begin{split}
E_V^{e-n} &= -\sum_{i=1}^{\eta}\sum_{p}\frac{Z_{p}}{|\bm{r_i}-\bm{R_p}|}|\psi(\bm{r_i})|^2, \\
E_V^{e-e} &= \sum_{i<j}\frac{1}{|\bm{r_i}-\bm{r_j}|}|\psi(\bm{r_i},\bm{r_j})|^2. 
\end{split}      
\end{equation}  
Probability distributions $|\psi(\bm{k_i})|^2$, $|\psi(\bm{r_i})|^2$, and $|\psi(\bm{r_i,r_j})|^2$ can be obtained 
by measuring the output of the (QFT applied) quantum circuit in the computational basis (Pauli $Z$ basis).
Note that, as pointed out by Chen et al., this method is not efficient in terms of sampling cost and is sensitive to discretization errors of grids~\cite{chan2023grid}.
According to their work, QPE should be used for measuring energy expectation values in future quantum calculations, 
but this naive method was adopted in the present study because of a lack of sufficient computational resources to simulate QPE on a classical computer.
The remaining issue in the first-quantized VQE is 
how to construct a variational quantum circuit that generates antisymmetric ansatz states.
In the following, we describe the design principle for constructing a variational quantum circuit to settle the above issue.

\subsection{Antisymmetrized variational quantum circuit}

To explain the design principles of circuits, 
we consider the one-dimensional two-electron system, 
which is the minimum model required to describe the antisymmetry of electrons.
In the following part of this section , we discuss two-electron systems, 
but we would like to note that the proposed method is applicable 
not only to two-electron systems but more generally to systems with larger numbers of electrons.

The overall circuit architecture is shown in Fig.~\ref{fig:overall_circuit}.
The proposed circuit consists of three parts: 
(1) seed state preparation, (2) one-body and two-body space variational circuits, 
and (3) measurement of energy expectations.
Since the measurement procedure of energy expectations has already been described, 
we here describe only the parts of the circuit that generate antisymmetric ansatz states.
In the first part of the circuit, some antisymmetric state, i.e., seed state, is prepared, 
at which the state vector of the circuit belongs to the antisymmetric subspace.
The subsequent one-body and two-body space variational circuits transform the state vector into the ground state 
while keeping it in the antisymmetric subspace.
Such a preparation of an antisymmetric state and antisymmetry-preserving transformations 
are the main design principle of the proposed circuit.
In the following, we describe each part of the circuit.

\subsubsection{Seed state preparation}
The simplest antisymmetric state of a two-electron system is the Greenberger-Horne-Zeilinger (GHZ) state, as follows:
\begin{equation}
\ket{\psi_{\mathrm{GHZ}}}=\frac{1}{\sqrt{2}}(\ket{0}_1\ket{1}_2-\ket{1}_1\ket{0}_2).
\end{equation}
This state can be generated by a series of an X (NOT) gate, an H (Hadamard) gate, and a CNOT (controlled NOT) gate as shown in Fig.~\ref{fig:overall_circuit}.
In this circuit, the spin coordinates $\ket{s^{(i)}}$ are antisymmetrized, 
meaning that this state is a singlet state 
in which the two spins are antiparallel $\downarrow_1\uparrow_2 - \uparrow_1\downarrow_2$ (0 is assumed to be $\downarrow$ and 1 to be $\uparrow$). 
This state is expressed as follows, including the spatial coordinates (the normalization constant $1/\sqrt{2}$ is omitted): 
\begin{equation}
\ket{\psi_{\mathrm{seed}}(\bm{r_1},\bm{r_2})} = \ket{\bm{0}}_1\ket{0}_1 \otimes \ket{\bm{0}}_2\ket{1}_2 - \ket{\bm{0}}_1\ket{1}_1 \otimes \ket{\bm{0}}_2\ket{0}_2, 
\end{equation}
where the spatial coordinate of each electron is $\ket{\bm{0}}=\ket{00\cdots0}$.
Obviously, this state is antisymmetric under the exchange of two electrons and can be used as a seed state.

For more-than-two-electron systems, seed states cannot be constructed in such a simple way.
As an example, for three-electron systems, 
the following Slater determinant consisting of the three states $\ket{00}$, $\ket{01}$, and $\ket{10}$ 
is one of the antisymmetric states (the normalization constant $1/\sqrt{3!}$ is omitted). 
\begin{equation}
\begin{split}
\ket{\psi_{00,01,10}}&=
\begin{vmatrix}
\ket{00}_1 & \ket{01}_1 & \ket{10}_1 \\
\ket{00}_2 & \ket{01}_2 & \ket{10}_2 \\
\ket{00}_3 & \ket{01}_3 & \ket{10}_3 \\
\end{vmatrix}\\
&=\ket{00}_1\ket{01}_2\ket{10}_3+\ket{01}_1\ket{10}_2\ket{00}_3\\
&\quad+\ket{10}_1\ket{00}_2\ket{01}_3-\ket{10}_1\ket{01}_2\ket{00}_3\\
&\quad-\ket{01}_1\ket{00}_2\ket{10}_3-\ket{00}_1\ket{10}_2\ket{01}_3.
\end{split}
\end{equation}
Previous research has proposed methods 
for constructing quantum circuits to prepare such a Slater determinant, 
though it is not as simple as that for the GHZ state.
Thanks to the work of Berry et al., 
an implementation of the Fisher-Yates shuffle~\cite{durstenfeld1964algorithm} on a quantum circuit has been provided 
and can generate the superpositions of an exponential number of permutation elements 
with polynomial computational complexity~\cite{berry2018improved}.
Following their method, it is possible to systematically generate seed states for more-than-two-electron systems.

Another way to prepare a seed state is a variational approach, 
which achieves a seed state using conventional variational circuits.
In this approach, the Hadamard test (swap test) can be employed as an objective function.
Letting $\mathrm{SWAP}_{ij}$ be the swap operator acting on the subspace of the $i$th and $j$th electrons, 
the output of the Hadamard test for the quantum state $\ket{\psi}$ becomes 
\begin{equation}
p_0 = \frac{1+\braket{\psi|\mathrm{SWAP}_{ij}|\psi}}{2}, p_1 = \frac{1-\braket{\psi|\mathrm{SWAP}_{ij}|\psi}}{2},
\end{equation}
where $p_0$ and $p_1$ are the measurement probabilities of the $\ket{0}$ and $\ket{1}$ of an ancilla qubit.
Suppose $\ket{\psi}$ is an antisymmetric wave function, $\mathrm{SWAP}_{ij}\ket{\psi} = -\ket{\psi}$; 
then the output of the Hadamard test becomes $p_0=0,p_1=1$ (for symmetric wave functions, $p_0=1,p_1=0$).
Therefore, an approximated antisymmetric state can be obtained 
by updating variational parameters to maximize the output of the Hadamard tests for all two-electron swap operations.
By repeatedly performing the Hadamard test on an approximated antisymmetric state, 
the pure antisymmetric states can eventually be distilled, 
which can then be used as a seed state.

\subsubsection{One-body space variational circuit}
The proposed variational quantum circuit consists of two components: 
a one-body space variational circuit and a two-body ($\eta$-body) space variational circuit.
We first explain the one-body space variational circuit.
A one-body space variational circuit consists of unitary operators 
that act on the subspace of each electron (one-body space).
Consider $\mathcal{U}^{(1)}(\theta) = \left[U^{(1)}(\theta)\right]^{\otimes\eta}$ as the one-body space variational circuit, 
where $U^{(1)}(\theta)$ is a unitary operator (variational circuit) acting on a one-body space. 
Since $U^{(1)}(\theta)$ transforms a state vector equally in each one-body space, 
$\mathcal{U}^{(1)}(\theta)$ does not destroy the antisymmetry of the ansatz state.
For two-electron systems, 
$\mathcal{U}^{(1)}(\theta)=U^{(1)}(\theta)\otimes U^{(1)}(\theta)$ acting on the seed state (GHZ state) $\ket{\psi_{\mathrm{seed}}(\bm{r_1}, \bm{r_2})}$ 
yields
\begin{equation}
\begin{split}
\ket{\psi(\bm{r_1}, \bm{r_2})} &= \mathcal{U}^{(1)}(\theta) \ket{\psi_{\mathrm{seed}}(\bm{r_1}, \bm{r_2})}\\
&=U^{(1)}(\theta) \ket{\bm{0}}_1\ket{0}_1 \otimes U^{(1)}(\theta) \ket{\bm{0}}_2\ket{1}_2 \\
&\quad- U^{(1)}(\theta) \ket{\bm{0}}_1\ket{1}_1 \otimes U^{(1)}(\theta) \ket{\bm{0}}_2\ket{0}_2\\
&=\ket{\alpha(\theta)}_1 \otimes \ket{\beta(\theta)}_2 - \ket{\beta(\theta)}_1 \otimes \ket{\alpha(\theta)}_2.
\end{split}
\end{equation}
As can be seen, $\mathcal{U}^{(1)}(\theta)$ transforms $\ket{\bm{0}}\ket{0},\ket{\bm{0}}\ket{1}$ 
into $\ket{\alpha(\theta)}=U^{(1)}(\theta)\ket{\bm{0}}\ket{0}$, $\ket{\beta(\theta)}=U^{(1)}(\theta)\ket{\bm{0}}\ket{1}$ in each one-body space, 
and antisymmetry is preserved under this transformation.
The resulting state can be expressed by a single Slater determinant consisting of $\ket{\alpha(\theta)}$ and $\ket{\beta(\theta)}$ as follows: 
\begin{equation}
\begin{split}
\ket{\psi_{\mathrm{SD}}(\bm{r_1}, \bm{r_2})} &=\ket{\alpha(\theta)}_1 \otimes \ket{\beta(\theta)}_2 - \ket{\beta(\theta)}_1 \otimes \ket{\alpha(\theta)}_2\\
&=
 \begin{vmatrix}
  \ket{\alpha(\theta)}_1 & \ket{\beta(\theta)}_1 \\
  \ket{\alpha(\theta)}_2 & \ket{\beta(\theta)}_2 
 \end{vmatrix}.    
\end{split} \label{eq:hf_state}
\end{equation}
This indicates that a one-body space variational circuit can only explore quantum states within the HF approximation.
By using the two-body ($\eta$-body) space variational circuit described in the following, 
we can explore quantum states beyond the HF approximation.

\subsubsection{Two-body space variational circuit}
The usual strategy to go beyond the HF approximation is based on configuration interaction (CI) theory, 
in which a many-body wave function is approximated by a linear combination, or superposition, of multiple Slater determinants.
As previously mentioned, a one-body space circuit generates an electronic state expressed by a single Slater determinant. 
Therefore, a superposition of multiple one-body space circuits is expected to generate 
a superposition of multiple Slater determinants, that is, an MC state.

For a two-electron system, 
consider two different operators $U_a\otimes U_a, U_b\otimes U_b$, 
where $U_a$ and $U_b$ are unitary operators acting on a one-body space .
Their superposition is expressed as 
\begin{equation}
U^{(2)}(\theta) = c_a(\theta) \cdot U_a \otimes U_a + c_b(\theta) \cdot U_b \otimes U_b ,
\end{equation}
where $c_a(\theta)$, $c_b(\theta)$ are superposition coefficients parametrized by $\theta$. 
Operator $U^{(2)}(\theta)$ acting on a single Slater determinant $\ket{\psi_{\mathrm{SD}}}$ yields 
\begin{equation}
\begin{split}
&U^{(2)}(\theta)\ket{\psi_{\mathrm{SD}}} \\
&= c_a(\theta)
\begin{vmatrix}
  U_a\ket{\alpha}_1 & U_a\ket{\beta}_1 \\
  U_a\ket{\alpha}_2 & U_a\ket{\beta}_2 \\
 \end{vmatrix}
+ c_{b}(\theta)
\begin{vmatrix}
  U_b\ket{\alpha}_1 & U_b\ket{\beta}_1 \\
  U_b\ket{\alpha}_2 & U_b\ket{\beta}_2 \\
 \end{vmatrix}\\
&=c_a(\theta)\ket{\psi^{a}_{\mathrm{SD}}}+c_b(\theta)\ket{\psi^{b}_{\mathrm{SD}}}. \label{eq:mc_state}
\end{split}
\end{equation}
As expected, a superposition of two different Slater determinants $\ket{\psi^{a}_{\mathrm{SD}}}$, $\ket{\psi^{b}_{\mathrm{SD}}}$, 
is generated by $U^{(2)}(\theta)$.

\begin{figure}[t]
  \centering
  \includegraphics[keepaspectratio, scale=0.3]{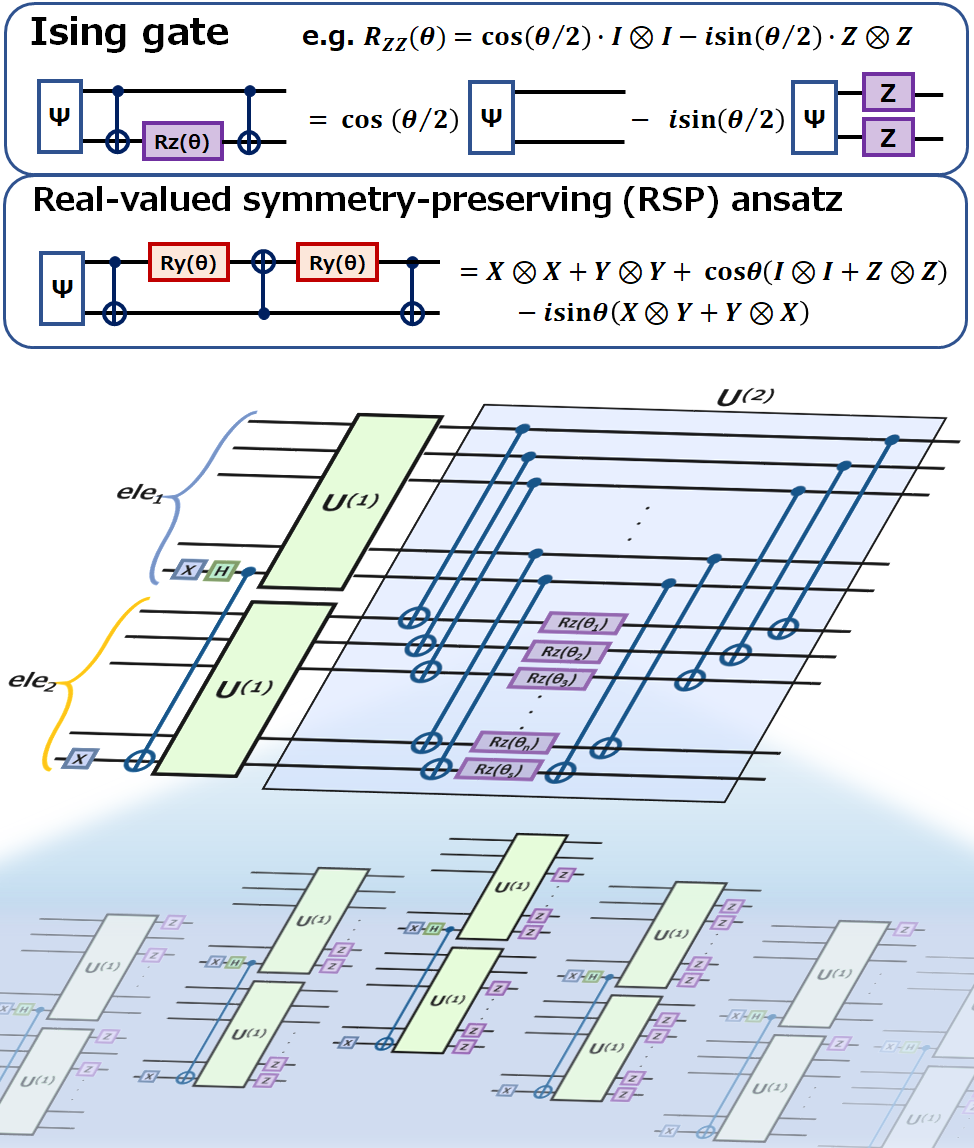}
  \caption{Implementation of $U^{(2)}(\theta)$. 
Top: implementations of Ising gate and real-valued symmetry-preserving (RSP) ansatz. 
Bottom: conceptual picture of generation of superposed HF states by an $R_{zz}(\theta)$ gate.}
  \label{fig:multi_conf}
\end{figure}

One of the simplest implementations of such an operator is the Ising gate~\cite{jones2003robust}. 
Ising gates such as $R_{xx}(\theta)$, $R_{yy}(\theta)$, and $R_{zz}(\theta)$ 
are represented by a superposition of $I \otimes I$ and $P \otimes P$ ($I$ is the identity operator, $P$ is the Pauli operator $X$, $Y$, $Z$); 
thus, they can be employed as $U^{(2)}(\theta)$. 
For example, $R_{zz}(\theta)$ shown in Fig.~\ref{fig:multi_conf} is represented by  
\begin{equation}
\begin{split}
R_{zz}(\theta) &=
  \begin{pmatrix}
    e^{-i\theta/2} & 0 & 0 & 0 \\
    0 & e^{i\theta/2} & 0 & 0 \\
    0 & 0 & e^{i\theta/2} & 0 \\
    0 & 0 & 0 & e^{-i\theta/2} \\
  \end{pmatrix}\\ 
&= \cos\frac{\theta}{2} \cdot I \otimes I -i\sin\frac{\theta}{2} \cdot Z \otimes Z.
\end{split}
\end{equation}
As can be seen from Fig.~\ref{fig:multi_conf}, the Ising gates act on a two-body space; 
thus, we refer to $U^{(2)}(\theta)$ as a two-body space variational circuit.
For $\eta$-electron systems, 
an $\eta$-body space variational circuit $U^{(\eta)}(\theta)$ is represented by $U^{(\eta)}(\theta) = \sum_{i} c_i(\theta) {U_i}^{\otimes \eta}$.
Such operators can be easily implemented using Ising gates with cascaded CNOT gates across the $\eta$-body space~\cite{whitfield2011simulation, kuhn2019accuracy}.

The Ising gate is not the only option as a two-body space variational circuit.
For example, the real-valued symmetry-preserving (RSP) ansatz~\cite{ibe2022calculating} shown in Fig.~\ref{fig:multi_conf} 
can also be used as a two-body space variational circuit. 
The RSP ansatz is represented by the superposition of the following operators: 
\begin{equation}
\begin{split}
U_{\mathrm{RSP}}(\theta) &= 
\begin{pmatrix}
  \cos \theta & 0 & 0 & -\sin \theta \\
  0 & 0 & 1 & 0 \\
  0 & 1 & 0 & 0 \\
  \sin \theta & 0 & 0 & \cos \theta \\
\end{pmatrix}\\
&=\frac{1}{2} \left[ X \otimes X + Y \otimes Y + \cos\theta( I \otimes I +  Z \otimes Z )\right. \\
&\qquad \left.- i\sin\theta(X \otimes Y + Y \otimes X) \right]. \label{eq:rsp}
\end{split}
\end{equation}
Here, terms $X\otimes Y, Y\otimes X$ appear
as tensor products of different operators $X$ and $Y$. 
Either of these terms alone destroys the antisymmetry of ansatz states, 
but pairs of them preserve antisymmetry as follows:
\begin{equation}
\begin{split}
&(X \otimes Y + Y \otimes X)\ket{\psi_{\mathrm{SD}}} \\
&=
\begin{vmatrix}
X\ket{\alpha}_1 & Y\ket{\beta}_1 \\
X\ket{\alpha}_2 & Y\ket{\beta}_2 \\
\end{vmatrix}
+
\begin{vmatrix}
Y\ket{\alpha}_1 & X\ket{\beta}_1 \\
Y\ket{\alpha}_2 & X\ket{\beta}_2 \\
\end{vmatrix}\\
&=\ket{\psi'_{\mathrm{SD}}}+\ket{\psi''_{\mathrm{SD}}}.
\end{split}
\end{equation}
In Eq.~\ref{eq:rsp}, the RSP ansatz is represented by a real-valued unitary operator, 
which is convenient for describing real-valued eigenstates of one-dimensional systems. 
Unfortunately, however, it is not obvious how to extend this ansatz to a more-than-two-body space. 
Although the Ising gates are considered a more suitable option for generalizing to a system with a larger number of electrons, 
in the present study, the RSP ansatz was employed for convenience.

As illustrated in Fig.~\ref{fig:multi_conf}, a single Slater determinant is split each time $U^{(2)}(\theta)$ acts on it.
Therefore, it is expected that an exponentially large number of Slater determinants will be generated
by repeatedly applying the two-body space variational circuit followed by the one-body space variational circuit.
However, such consecutive application of two-body variational circuits does not increase the number of superposed Slater determinants as one might expect. 
This is because a product of multiple Pauli operators reduces to a single Pauli operator 
due to the commutation relation among Pauli operators ($P_i^2=I,P_1P_2=iP_3$).
This reduction of Pauli operators can be prevented by using the alternating layered structure of one-body and two-body space variational circuits as shown in Fig.~\ref{fig:overall_circuit}.
If the one-body space variational circuit is noncommutative with the Pauli operators, 
then the reduction of Pauli operators between two-body space variational circuits is prevented.
Therefore, by repeatedly applying one-body and two-body space variational circuits alternately to a seed state,
the number of superposed Slater determinants can be increased exponentially, 
and so is a systematic approach to approximating the exact ground state.

During the optimization process of the VQE framework, 
variational parameters in one-body and two-body circuits are simultaneously optimized 
so as to minimize the energy expectation values of an ansatz state.
As can be seen from Eqs.~\ref{eq:hf_state}--\ref{eq:mc_state}, 
optimizing one-body and two-body circuits corresponds to optimizing 
the electronic states composing each Slater determinant (electronic orbitals) and the superposition coefficients of the Slater determinants (CI coefficients), respectively. 
Therefore, this framework can be regarded as the multi-configuration self-consistent field (MCSCF) method, 
which is one of the post-Hartree-Fock \textit{ab initio} quantum chemistry methods. 
A notable advantage of the proposed method is 
its ability to perform MCSCF calculations based on an exponential number of electron configurations 
with polynomial computational complexity.

\subsection{Settings for Numerical Simulations}
To verify the validity of our method, 
we performed the VQE calculation with the proposed circuit to obtain the ground state of a 1D-$\rm{H_2}$ system. 
The Hamiltonian of this system is expressed as follows: 
\begin{equation}
\begin{split}
H&=\sum_{i=1}^{2}\left[-\frac{1}{2}\frac{d^2}{dr_i^2}-\sum_{p=1}^{2}\frac{1}{\left|r_i-R_p+\epsilon\right|}\right]\\
&\qquad +\frac{1}{|r_1-r_2+\epsilon|}+\frac{1}{|R_1-R_2|}.  \label{eq:hydrogen}
\end{split}
\end{equation}
where $r_1, r_2$ and $R_1, R_2$ are the positions of electrons and protons, respectively.
To avoid zero division in the Coulomb interaction terms, 
a small real value $\epsilon$ is added to the denominators of those terms (soft Coulomb potential).

\begin{figure}[t]
  \centering
  \includegraphics[keepaspectratio, scale=0.35]{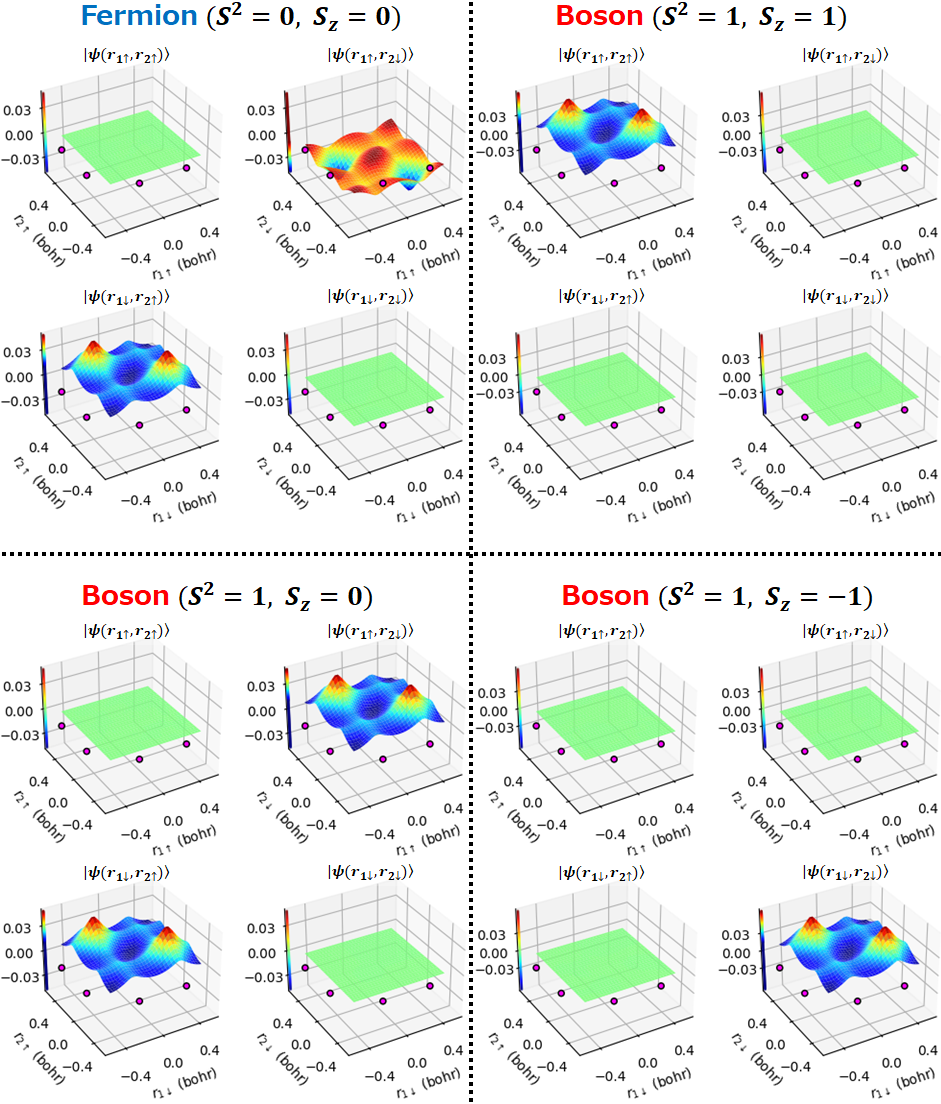}
  \caption{Probability amplitudes $\ket{\psi(r_1,r_2)}$ of degenerated ground states (one fermionic state ($S^2=0, S_z=0$) and three bosonic states ($S^2=1, S_z=0,\pm1$)) at interatomic distance $|R_1-R_2| = $ 0.5 bohrs ($R_1, R_2$ are shown by pink dots). 
  Each many-body wave function is shown in the four subspaces corresponding to the spin configurations $\uparrow_1\uparrow_2, \uparrow_1\downarrow_2, \downarrow_1\uparrow_2, \downarrow_1\downarrow_2$.
  Note that the correspondence between colors and values is different for each plot.}
  \label{fig:exact}
\end{figure}

Note here that the ground states of fermions and bosons are completely degenerate.
Figure~\ref{fig:exact}  shows four degenerate ground states obtained by the exact diagonalization of Eq.~\ref{eq:hydrogen} at the interatomic distance $|R_1-R_2| = $ 0.5 bohrs.
As shown in Fig.~\ref{fig:exact}, the ground states of the fermions are the singlet state ($S^2=0, S_z=0$) whose spin symmetry is antisymmetric and spatial symmetry is symmetric. 
On the other hand, the ground states of the bosons are the triplet state ($S^2=1, S_z=0,\pm1$) whose spin and spatial symmetry are symmetric. 
Since the Hamiltonian of the system depends only on the spatial coordinate, 
there is no energy difference between fermionic and bosonic ground states whose spatial symmetries are the same. 
Therefore, without taking into account the symmetry of the variational quantum circuit, 
the quantum states obtained by the VQE can be a mixture of fermionic and bosonic states, 
even though an accurate energy value is obtained.
We demonstrate such convergence to a symmetry-neglected state in the numerical simulations.

The following describes the calculation conditions of the system and the quantum circuits used for numerical simulations.
Six qubits  (5 qubits for spatial coordinate and 1 qubit for spin coordinate) represent the one-body space, 
and thus 12 qubits were used for the two-electron system.
Spatial coordinates of electrons $r_1, r_2$ ranged from -0.5 bohrs to 0.5 bohrs and the grid width $\varDelta r$ was $1/2^{5}$ bohrs = 0.03125 bohrs.
$\epsilon$ was set to $\varDelta r/2$.
The spatial grid employed in this study was limited to a very coarse one, due to the computational resources of our classical computer, 
but the realization of the FTQC is expected to allow calculations with a grid fine enough to reproduce the electronic structure of real materials.
For variational circuits, 
the hardware efficient (HE) ansatz~\cite{kandala2017hardware} and the real-valued symmetry preserving ansatz (RSP)~\cite{ibe2022calculating} were employed.
The HE ansatz consists of Ry gates and CNOT gates~\cite{shee2022qubit}, 
which in the present study had 6 layers.

\begin{figure}[t]
  \subfloat[]{\includegraphics[keepaspectratio, scale=0.45]{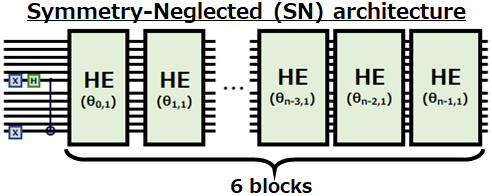}\label{fig:circ_sn}}\\
  \subfloat[]{\includegraphics[keepaspectratio, scale=0.45]{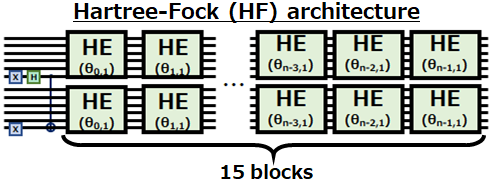}\label{fig:circ_hf}}\\
  \subfloat[]{\includegraphics[keepaspectratio, scale=0.45]{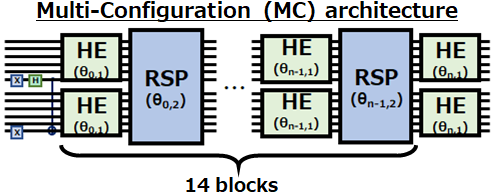}\label{fig:circ_mc}}\\
  \caption{Circuit architectures employed by VQE. (a) Symmetry-neglected  (SN) architecture; (b) Hartree-Fock (HF) architecture; (c) Multi-configuration (MC) architecture.
  HE in the figure denotes the hardware efficient ansatz with 6 layers and RSP denotes the real-valued symmetry preserving ansatz.}
  \label{fig:circ}
\end{figure}

In order to demonstrate the effects of the antisymmetrization and the multi-configuration, 
we implemented the VQE with three different circuits.
The architectures of these circuits are shown in Fig.~\ref{fig:circ}.
The first architecture is the symmetry-neglected (SN) architecture shown in Fig.~\ref{fig:circ}(a), 
consisting of consecutive HE ansatz blocks acting on the two-body space. 
In this architecture, the antisymmetry of the seed state is not considered to be preserved, 
and thus it is expected that the VQE will yield a neither antisymmetric (fermionic) nor symmetric (bosonic) state.
The number of blocks of the HE ansatz was 6.
The second architecture is the Hartree-Fock (HF) architecture shown in Fig.~\ref{fig:circ}(b), 
consisting of consecutive one-body space variational circuits. 
In this architecture, the antisymmetry of the seed state is preserved, 
but the VQE yields a ground state within the HF approximation.
The third architecture is the multi-configuration (MC) architecture shown in Fig.~\ref{fig:circ}(c), 
consisting of alternately layered one-body and two-body space variational circuits. 
In this architecture, the VQE can achieve a ground state beyond the HF approximation, 
and its energy is expected to be lower than that obtained with the HF architecture.

To confirm this energy stabilization, potential energy curves were calculated for the HF and MC architectures,  
at 16 points of interatomic distance $|R_1-R_2|$ ranging from $\varDelta r$ to $16\varDelta r$ = 0.5 bohrs. 
The number of one-body space variational circuits was 15 for both architectures 
and that of two-body space variational circuits was 14 for the MC architecture. 
For the optimization of variational parameters, 
the steepest descent method with the Adam optimizer~\cite{kingma2014adam} was employed 
and the number of optimization steps was set to 10000.
All calculations for quantum circuits were performed by a noiseless state vector simulator.

\section{Result and discussion}

\begin{figure*}[!t]
  \centering
  \subfloat[]{\includegraphics[keepaspectratio, scale=0.45]{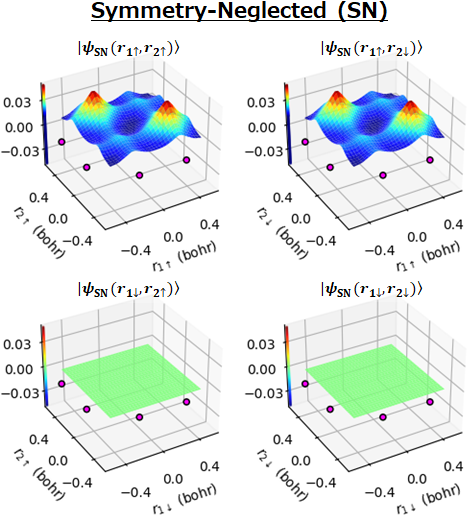}\label{fig:wf_sn}}\hspace{2.2mm}
  \subfloat[]{\includegraphics[keepaspectratio, scale=0.45]{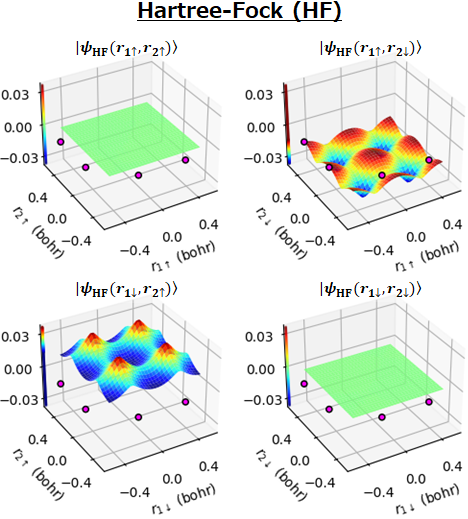}\label{fig:wf_hf}}\hspace{2.2mm}
  \subfloat[]{\includegraphics[keepaspectratio, scale=0.45]{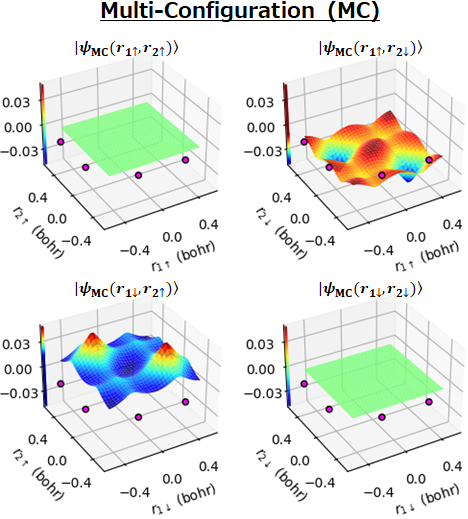}\label{fig:wf_mc}}
  \caption{Many-body wave functions obtained by the VQE 
  with (a) the SN architecture, (b) the HF architecture and (c) the MC architecture 
  at the interatomic distance $|R_1-R_2| = $ 0.5 bohrs ($R_1, R_2$ are shown by pink dots).}
  \label{fig:wave_func}
\end{figure*}

\subsection{Results of the VQE Calculations}

Figure~\ref{fig:wave_func} shows the many-body wave functions obtained by the VQE with the different architectures at an interatomic distance $|R_1-R_2| = $ 0.5 bohrs.
In the case of the wave function obtained with the SN architecture shown in Fig.~\ref{fig:wave_func}(a), 
the spatial coordinate is symmetric under the exchange of two electrons, 
but its spin coordinate is neither antisymmetric nor symmetric ($\uparrow_1\uparrow_2+\uparrow_1\downarrow_2\neq\uparrow_1\uparrow_2+\downarrow_1\uparrow_2$).
This indicates that unless the symmetry for the variational quantum circuit is not considered, 
the resulting state of the VQE can converge to such a symmetry-neglected state. 
This is the notable difference from the second-quantized VQE.
In contrast, as shown in Figs.~\ref{fig:wave_func}(b)(c), 
the wave functions obtained with the HF and MC architectures are the antisymmetric singlet states, as expected. 
The shape of the wave function obtained with the MC architecture well reproduces that of the exact fermionic ground state shown in Fig.~\ref{fig:exact}, 
while that with the HF architecture is apparently different from it.
As will be described later, this difference in shape of the wave functions indicates the difference in representability of the quantum states between the HF approximation and CI theory.

\begin{figure}[bthp]
  \subfloat[]{\includegraphics[keepaspectratio, scale=0.52]{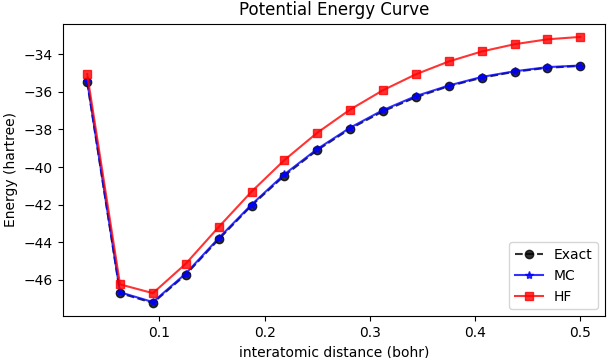}\label{fig:pec}}\\
  \subfloat[]{\includegraphics[keepaspectratio, scale=0.52]{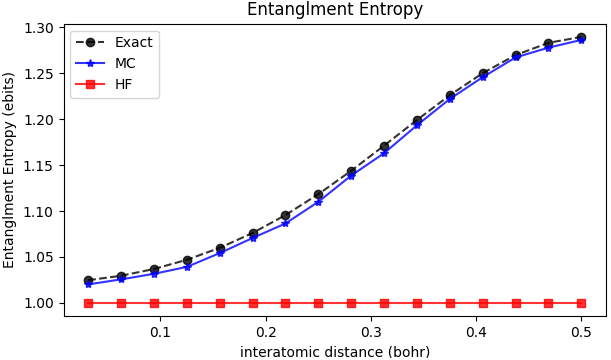}\label{fig:ee}}\\
  \caption{(a) Potential energy curves and (b) entanglement entropy 
  obtained by exact diagonalization (black dashed lines) and VQE with the MC architecture (blue solid lines) and the HF architecture (red solid lines).
  The calculated potential energy curves are significantly different from those of the three-dimensional hydrogen molecular system, 
  which is attributed to the difference in dimensions between the systems~\cite{hirai2022molecular}.}
  \label{fig:pec_ee}
\end{figure}

The difference between them is also reflected in their ground state energies.
Figure~\ref{fig:pec_ee}(a) shows the potential energy curves obtained with the HF and MC architectures.
It can be seen that 
the ground state energies obtained with the MC architecture reproduce the results of the exact diagonalization well, 
whereas those obtained with the HF architecture converge to higher values.
This demonstrates the energy stabilization due to the multi-configuration character, in other words, the effect of electron correlation, 
which is the pillar of the quantum chemistry theories~\cite{jensen2017introduction, helgaker2013molecular}. 

\subsection{Analysis of Many-Body States based on Quantum Information Theory}

Although the numerical experiments in this section confirmed that the proposed method can reproduce the exact fermionic ground state well, 
the obtained many-body wave function provides little insight into the microscopic electronic structure.
The electronic structure is mostly understood from the electron orbital picture; 
however, since our method is based on a real-space basis, 
the orbital picture of the obtained state is not obvious.
Furthermore, a lack of an orbital picture makes it impossible to quantitatively evaluate the multi-configuration character of the obtained state.
In order to get a clear orbital picture and evaluate the multi-configuration character of obtained states, 
we analyzed many-body quantum states from the perspective of quantum information theory.
Quantum information has been attracting increasing attention in recent years 
as a key to understanding electron dynamics in chemical reactions and strongly correlated materials
~\cite{molina2015quantum, brandejs2019quantum, esquivel2011quantum, rissler2006measuring, boguslawski2013orbital, lin2013spatial, soh2019long, zhu2020entanglement, lanata2014principle, chen2019incoherent}.

We will first quantify the multi-configuration character of many-body states.
To evaluate the multi-configuration character, 
we employed the entanglement entropy, which is typically employed to measure the degree of entanglement between subsystems.
We shall now briefly explain the definition of entanglement entropy.
A many-body wave function of the two-electron system can be decomposed into multiple product states by the Schmidt decomposition, as follows: 
\begin{equation}
\ket{\psi(r_1,r_2)} = \sum_i \lambda_i \ket{\mu_i(r_1)} \otimes \ket{\chi_i(r_2)}, \label{eq:Schmidt}
\end{equation}
where $\lambda_i$ are the Schmidt coefficients (superposition coefficients of product states) 
and $\ket{\mu_i(r_1)}, \ket{\chi_i(r_2)}$ are the Schmidt basis on subsystems.
The entanglement entropy $S$ is defined by the Shannon entropy of the probability distribution $|\lambda_i|^2$ as
\begin{equation}
S = -\sum_i |\lambda_i|^2 \log_2|\lambda_i|^2. \label{eq:ee}
\end{equation}
Considering the fact that the Shannon entropy of a normal distribution is proportional to the logarithm of its variance, 
the larger the variance of probability distribution $|\lambda_i|^2$, that is, the more product states are superposed, the larger the entanglement entropy.
Since the entanglement entropy of 
a single Slater determinant $\ket{\psi_{\rm{HF}}}=1/\sqrt2(\ket{\alpha}_1 \otimes \ket{\beta}_2 - \ket{\beta}_1 \otimes \ket{\alpha}_2)$ is 1, 
that of an MC state is expected to be larger than 1.

Figure~\ref{fig:pec_ee}(b) shows the entanglement entropies calculated for the ground states obtained by the exact diagonalization and the MC architecture.
As expected, the entanglement entropy of the MC state is larger than 1 and increases as interatomic distance increases, 
which is in good agreement with the exact result.
This behavior of entanglement entropy indicates that 
the ground state becomes more multi-configurational as the interatomic distance approaches the dissociation limit, 
which is a well-known fact in the field of conventional quantum chemistry~\cite{jensen2017introduction, helgaker2013molecular}.

We can now extract the electron orbital picture from the many-body state represented in real space. 
From the expression of the Schmidt decomposition in Eq.~\ref{eq:Schmidt}, 
Schmidt basis $\ket{\mu_i(r_1)},\ket{\chi_i(r_2)}$ can be regarded as one-body wave functions, that is, the electron orbitals of the electrons, 
and the Schmidt coefficients as their contributions to the many-body state.
Therefore, the Schmidt basis and the Schmidt coefficients provide insight into 
the electron orbital picture of many-body wave functions.

\begin{figure}[t]
  \centering
  \includegraphics[keepaspectratio, scale=0.5]{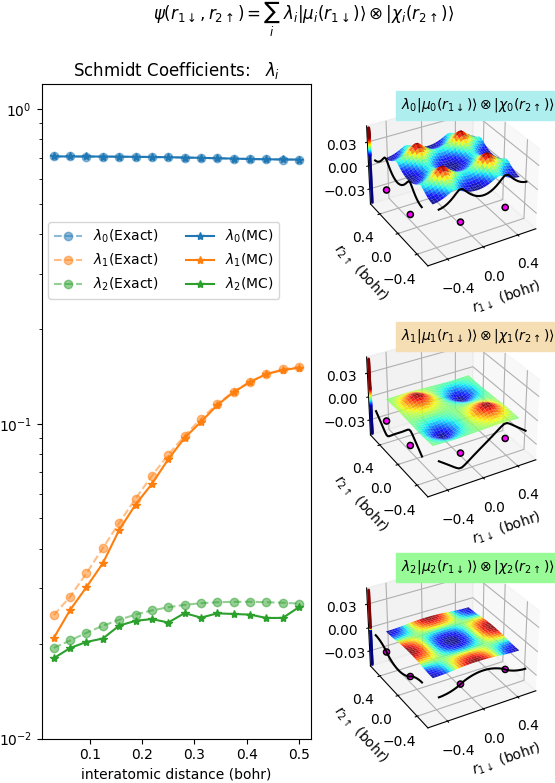}
  \caption{Results of the Schmidt decomposition of $\ket{\psi(r_{1\downarrow},r_{2\uparrow})}$. 
  Left: First three Schmidt coefficients $\lambda_0, \lambda_1, \lambda_2$ at each interatomic distance. 
  Right: Resulting product-state wave function at interatomic distance $|R_1-R_2|$ = 0.5 bohrs ($R_1, R_2$ are shown by pink dots), 
  and corresponding one-electron orbitals $\ket{\mu_i(r_{1\downarrow})}, \ket{\chi_i(r_{2\uparrow})}$ (black lines). The one-electron orbitals are scaled for visibility.}
  \label{fig:svd}
\end{figure}

Figure~\ref{fig:svd} shows the results of the Schmidt decomposition performed for one anti-parallel spin subspace $\ket{\psi(r_{1\downarrow},r_{2\uparrow})}$ 
of the exact ground state (upper left part of Fig.~\ref{fig:exact}) and the MC state (Fig.~\ref{fig:wave_func}(c)).
The left side of the figure shows the first three Schmidt coefficients $\lambda_0, \lambda_1, \lambda_2$ of each interatomic distance.
As shown, 
the contribution of the zeroth product state $\lambda_0$ is nearly constant at all interatomic distances, 
whereas that of the first product state $\lambda_1$ increases as interatomic distance increases, 
which is consistent with the behavior of the entanglement entropy shown in Fig.~\ref{fig:pec_ee}(b).
This suggests that the configuration interaction between the zeroth and first product states, that is, their superposition, 
explains the energy stabilization due to the multi-configuration character.
The right side of the figure shows the product states constituting the MC state at the interatomic distance $|R_1-R_2| = $ 0.5 bohrs.
The electron orbitals constituting each product state are indicated by black lines in each figure.
Using these electron orbitals, we will illustrate the energy stabilization due to the multi-configuration character in the following discussion.

The electron orbitals constituting the zeroth and first product states have peaks at the proton positions $R_1, R_2$ (pink dots in figure), 
and these orbitals correspond to bonding and antibonding orbitals.
Following the typical explanation of the linear combination of atomic orbitals (LCAO) approximation, 
let $\ket{p_i(r)}$ be the electron orbitals  distributed around the $i$th proton. 
Then the bonding and antibonding orbitals are represented as 
$\ket{\psi_{\mathrm{\sigma}}(r)}=\ket{p_1(r)}+\ket{p_2(r)}$ and $\ket{\psi_{\mathrm{\sigma^{*}}}(r)}=\ket{p_1(r)}-\ket{p_2(r)}$, respectively.
The bonding state of two electrons $\ket{\psi_{\mathrm{\sigma\sigma}}(r_1,r_2)}$ 
is expressed as the tensor product of the bonding orbitals $\ket{\psi_{\mathrm{\sigma}}(r)}$ as follows: 
\begin{equation}
  \begin{split}
  &\ket{\psi_{\mathrm{\sigma\sigma}}(r_1,r_2)}\\
  &=\ket{\psi_{\mathrm{\sigma}}(r_1)}\otimes\ket{\psi_{\mathrm{\sigma}}(r_2)}  \\
  &= \ket{p_1(r_1)} \otimes \ket{p_2(r_2)} + \ket{p_2(r_1)} \otimes \ket{p_1(r_2)}\\
  &\qquad + \ket{p_1(r_1)} \otimes \ket{p_1(r_2)} + \ket{p_2(r_1)} \otimes \ket{p_2(r_2)},
  \end{split}
\end{equation}
The four terms in the above expression correspond to the four peaks 
in the zeroth product state and in the HF ground state shown in Fig.~\ref{fig:wave_func}(b); thus,
\begin{equation}
\ket{\psi_{\mathrm{\sigma\sigma}}(r_1,r_2)} \sim \ket{\mu_0(r_1)}\otimes\ket{\chi_0(r_2)} \sim \ket{\psi_{\mathrm{HF}}(r_1,r_2)}.    
\end{equation}
The antibonding state, or two-electron excited state, $\ket{\psi_{\mathrm{\sigma^{*}\sigma^{*}}}(r_1,r_2)}$ 
is expressed as the tensor product of the antibonding orbitals $\ket{\psi_{\mathrm{\sigma^{*}}}(r)}$ as follows: 
\begin{equation}
  \begin{split}
  &\ket{\psi_{\mathrm{\sigma^{*}\sigma^{*}}}(r_1,r_2)}\\
  &=-\ket{\psi_{\mathrm{\sigma^{*}}}(r_1)}\otimes\ket{\psi_{\mathrm{\sigma^{*}}}(r_2)} \\
  &= \ket{p_1(r_1)} \otimes \ket{p_2(r_2)} + \ket{p_2(r_1)} \otimes \ket{p_1(r_2)}\\
  &\qquad - \ket{p_1(r_1)} \otimes \ket{p_1(r_2)} - \ket{p_2(r_1)} \otimes \ket{p_2(r_2)},
  \end{split}
\end{equation}
where $-1$ is taken as a global phase.
The four terms in the above expression correspond to the four peaks 
in the first product state; thus, 
\begin{equation}
\ket{\psi_{\mathrm{\sigma^{*}\sigma^{*}}}(r_1,r_2)} \sim \ket{\mu_1(r_1)}\otimes\ket{\chi_1(r_2)}.
\end{equation}
Superposing bonding and antibonding states equally yields 
\begin{equation}
  \begin{split}
  &\ket{\psi_{\mathrm{\sigma\sigma+\sigma^{*}\sigma^{*}}}(r_1,r_2)}\\
  &=\frac{1}{2}\ket{\psi_{\mathrm{\sigma\sigma}}(r_1,r_2)} + \frac{1}{2}\ket{\psi_{\mathrm{\sigma^{*}\sigma^{*}}}(r_1,r_2)}\\
  &= \ket{p_1(r_1)} \otimes \ket{p_2(r_2)} + \ket{p_2(r_1)} \otimes \ket{p_1(r_2)}.
  \end{split}
\end{equation}
It can be seen that energetically unfavorable electron configurations $\ket{p_1(r_1)} \otimes \ket{p_1(r_2)}$, $\ket{p_2(r_1)} \otimes \ket{p_2(r_2)}$, 
where both electrons are distributed around the same proton, are cancelled out and a lower energy state is achieved. 
The two terms in the above expression correspond to the two peaks in the MC state $\ket{\psi_{\mathrm{MC}}(r_1,r_2)}$ shown in Fig.~\ref{fig:wave_func}(c); thus, 
\begin{equation}
\ket{\psi_{\mathrm{\sigma\sigma+\sigma^{*}\sigma^{*}}}(r_1,r_2)} \sim \ket{\psi_{\mathrm{MC}}(r_1,r_2)}.   
\end{equation}
This state represents the electron configuration where electrons are distributed around different protons to avoid each other, 
reminiscent of correlated, or entangled, electrons.
To summarize, a multi-configuration state is inherently an entangled state consisting of multiple product states, 
and electron correlation can be interpreted as quantum entanglement in the electron system.

Although this discussion on electron orbitals may be just a textbook subject~\cite{jensen2017introduction, helgaker2013molecular}, 
it is worth noting that such an orbital picture can be extracted from a real-space represented by many-body states by using the Schmidt decomposition.
In addition, the minor orbital contributions, other than bonding and antibonding orbitals, can be considered in our method.
Indeed, as shown in the left part of Fig.~\ref{fig:svd}, the exact ground state and the MC state include 
the contribution of the second product state $\lambda_2\ket{\mu_2(r_{1\downarrow})}\otimes\ket{\chi_2(r_{2\uparrow})}$ 
whose electron orbitals have nodes at proton positions and are distributed in the interstitial region.
Since a real-space basis is a complete basis set of a discrete space, 
it can represent both localized atomic orbitals and such delocalized orbitals, 
freeing us from the need for careful selection of basis functions in typical quantum chemical calculations.
From this point, a real-space basis is an attractive option for the quantum simulation in the FTQC era.

The results shown in this section demonstrate one possible framework of quantum chemistry in the FTQC era, 
where we understand electronic structures based on electron orbitals derived from many-body wave functions represented in real space.
In this study, many-body wave functions were analyzed by a classical computer; however, 
since it is almost impossible to obtain a state vector of even a quantum circuit of a few dozen qubits, 
this analysis must be performed on a quantum computer.
Fortunately, several methods have been proposed to perform singular value decomposition on quantum computers~\cite{rebentrost2018quantum, bravo2020quantum, wang2021variational}, 
which may allow such analysis to be performed on a quantum computer with reasonable computational resources.

The problem concerned in our method is the barren plateaus~\cite{mcclean2018barren, cerezo2021cost, wang2021noise} in the optimization process of the variational state.
In fact, the disappearance of gradients of variational parameters was observed during the optimization process, 
which required a large number of optimization steps.
Such barren plateaus are the most serious problem in most variational quantum algorithms.
Nevertheless, as mentioned in the Introduction, 
since by using the QPE, the true ground state can be distilled from the approximated ground state prepared by the VQE, 
the incomplete convergence of the VQE due to the barren plateaus will be tolerated to some extent. 
The complementary combination of the VQE and the QPE will allow the state preparation for first-quantized methods.

\section{Summary and Conclusion}
In this paper, we propose a state preparation method for the first-quantized quantum simulations based on the real-space representation.
We employ a variational method for preparing ground states, 
and provide a design principle for constructing antisymmetrized variational quantum circuits.
Our proposed circuits are capable of generating a superposition of exponentially large number of Slater determinants, that is, MC states with polynomial numbers of quantum gates. 
We performed VQE calculations for a 1D-$\rm{H_2}$ system 
and confirmed that the proposed circuit reproduces well the exact fermionic ground state.
In addition to performing VQE, we quantitatively evaluated the multi-configuration character of many-body states by using the entanglement entropy between two electrons, 
and extracted the electronic orbital picture from many-body states represented in real space by the Schmidt decomposition.

Quantum computers, as demonstrated in this study, have great potential to simulate many-body electron systems and shed light on their quantum information.
We believe that our proposed method will contribute to realizing the first-quantized simulation for electron dynamics 
and will bring a deeper understanding of electron systems to materials science in the FTQC era.

\nocite{*}
\bibliographystyle{apsrev4-2}
\bibliography{main_ref.bib} 

\begin{thebibliography}{59}%
\makeatletter
\providecommand \@ifxundefined [1]{%
 \@ifx{#1\undefined}
}%
\providecommand \@ifnum [1]{%
 \ifnum #1\expandafter \@firstoftwo
 \else \expandafter \@secondoftwo
 \fi
}%
\providecommand \@ifx [1]{%
 \ifx #1\expandafter \@firstoftwo
 \else \expandafter \@secondoftwo
 \fi
}%
\providecommand \natexlab [1]{#1}%
\providecommand \enquote  [1]{``#1''}%
\providecommand \bibnamefont  [1]{#1}%
\providecommand \bibfnamefont [1]{#1}%
\providecommand \citenamefont [1]{#1}%
\providecommand \href@noop [0]{\@secondoftwo}%
\providecommand \href [0]{\begingroup \@sanitize@url \@href}%
\providecommand \@href[1]{\@@startlink{#1}\@@href}%
\providecommand \@@href[1]{\endgroup#1\@@endlink}%
\providecommand \@sanitize@url [0]{\catcode `\\12\catcode `\$12\catcode
  `\&12\catcode `\#12\catcode `\^12\catcode `\_12\catcode `\%12\relax}%
\providecommand \@@startlink[1]{}%
\providecommand \@@endlink[0]{}%
\providecommand \url  [0]{\begingroup\@sanitize@url \@url }%
\providecommand \@url [1]{\endgroup\@href {#1}{\urlprefix }}%
\providecommand \urlprefix  [0]{URL }%
\providecommand \Eprint [0]{\href }%
\providecommand \doibase [0]{https://doi.org/}%
\providecommand \selectlanguage [0]{\@gobble}%
\providecommand \bibinfo  [0]{\@secondoftwo}%
\providecommand \bibfield  [0]{\@secondoftwo}%
\providecommand \translation [1]{[#1]}%
\providecommand \BibitemOpen [0]{}%
\providecommand \bibitemStop [0]{}%
\providecommand \bibitemNoStop [0]{.\EOS\space}%
\providecommand \EOS [0]{\spacefactor3000\relax}%
\providecommand \BibitemShut  [1]{\csname bibitem#1\endcsname}%
\let\auto@bib@innerbib\@empty
\bibitem [{\citenamefont {McArdle}\ \emph {et~al.}(2020)\citenamefont
  {McArdle}, \citenamefont {Endo}, \citenamefont {Aspuru-Guzik}, \citenamefont
  {Benjamin},\ and\ \citenamefont {Yuan}}]{mcardle2020quantum}%
  \BibitemOpen
  \bibfield  {author} {\bibinfo {author} {\bibfnamefont {S.}~\bibnamefont
  {McArdle}}, \bibinfo {author} {\bibfnamefont {S.}~\bibnamefont {Endo}},
  \bibinfo {author} {\bibfnamefont {A.}~\bibnamefont {Aspuru-Guzik}}, \bibinfo
  {author} {\bibfnamefont {S.~C.}\ \bibnamefont {Benjamin}},\ and\ \bibinfo
  {author} {\bibfnamefont {X.}~\bibnamefont {Yuan}},\ }\bibfield  {title}
  {\bibinfo {title} {Quantum computational chemistry},\ }\href@noop {}
  {\bibfield  {journal} {\bibinfo  {journal} {Reviews of Modern Physics}\
  }\textbf {\bibinfo {volume} {92}},\ \bibinfo {pages} {015003} (\bibinfo
  {year} {2020})}\BibitemShut {NoStop}%
\bibitem [{\citenamefont {Bauer}\ \emph {et~al.}(2020)\citenamefont {Bauer},
  \citenamefont {Bravyi}, \citenamefont {Motta},\ and\ \citenamefont
  {Chan}}]{bauer2020quantum}%
  \BibitemOpen
  \bibfield  {author} {\bibinfo {author} {\bibfnamefont {B.}~\bibnamefont
  {Bauer}}, \bibinfo {author} {\bibfnamefont {S.}~\bibnamefont {Bravyi}},
  \bibinfo {author} {\bibfnamefont {M.}~\bibnamefont {Motta}},\ and\ \bibinfo
  {author} {\bibfnamefont {G.~K.-L.}\ \bibnamefont {Chan}},\ }\bibfield
  {title} {\bibinfo {title} {Quantum algorithms for quantum chemistry and
  quantum materials science},\ }\href@noop {} {\bibfield  {journal} {\bibinfo
  {journal} {Chemical Reviews}\ }\textbf {\bibinfo {volume} {120}},\ \bibinfo
  {pages} {12685--12717} (\bibinfo {year} {2020})}\BibitemShut {NoStop}%
\bibitem [{\citenamefont {Ma}\ \emph {et~al.}(2020)\citenamefont {Ma},
  \citenamefont {Govoni},\ and\ \citenamefont {Galli}}]{ma2020quantum}%
  \BibitemOpen
  \bibfield  {author} {\bibinfo {author} {\bibfnamefont {H.}~\bibnamefont
  {Ma}}, \bibinfo {author} {\bibfnamefont {M.}~\bibnamefont {Govoni}},\ and\
  \bibinfo {author} {\bibfnamefont {G.}~\bibnamefont {Galli}},\ }\bibfield
  {title} {\bibinfo {title} {Quantum simulations of materials on near-term
  quantum computers},\ }\href@noop {} {\bibfield  {journal} {\bibinfo
  {journal} {npj Computational Materials}\ }\textbf {\bibinfo {volume} {6}},\
  \bibinfo {pages} {85} (\bibinfo {year} {2020})}\BibitemShut {NoStop}%
\bibitem [{\citenamefont {Nam}\ \emph {et~al.}(2020)\citenamefont {Nam},
  \citenamefont {Chen}, \citenamefont {Pisenti}, \citenamefont {Wright},
  \citenamefont {Delaney}, \citenamefont {Maslov}, \citenamefont {Brown},
  \citenamefont {Allen}, \citenamefont {Amini}, \citenamefont {Apisdorf} \emph
  {et~al.}}]{nam2020ground}%
  \BibitemOpen
  \bibfield  {author} {\bibinfo {author} {\bibfnamefont {Y.}~\bibnamefont
  {Nam}}, \bibinfo {author} {\bibfnamefont {J.-S.}\ \bibnamefont {Chen}},
  \bibinfo {author} {\bibfnamefont {N.~C.}\ \bibnamefont {Pisenti}}, \bibinfo
  {author} {\bibfnamefont {K.}~\bibnamefont {Wright}}, \bibinfo {author}
  {\bibfnamefont {C.}~\bibnamefont {Delaney}}, \bibinfo {author} {\bibfnamefont
  {D.}~\bibnamefont {Maslov}}, \bibinfo {author} {\bibfnamefont {K.~R.}\
  \bibnamefont {Brown}}, \bibinfo {author} {\bibfnamefont {S.}~\bibnamefont
  {Allen}}, \bibinfo {author} {\bibfnamefont {J.~M.}\ \bibnamefont {Amini}},
  \bibinfo {author} {\bibfnamefont {J.}~\bibnamefont {Apisdorf}}, \emph
  {et~al.},\ }\bibfield  {title} {\bibinfo {title} {Ground-state energy
  estimation of the water molecule on a trapped-ion quantum computer},\
  }\href@noop {} {\bibfield  {journal} {\bibinfo  {journal} {npj Quantum
  Information}\ }\textbf {\bibinfo {volume} {6}},\ \bibinfo {pages} {33}
  (\bibinfo {year} {2020})}\BibitemShut {NoStop}%
\bibitem [{\citenamefont {Shirai}\ \emph {et~al.}(2022)\citenamefont {Shirai},
  \citenamefont {Horiba},\ and\ \citenamefont {Hirai}}]{shirai2022calculation}%
  \BibitemOpen
  \bibfield  {author} {\bibinfo {author} {\bibfnamefont {S.}~\bibnamefont
  {Shirai}}, \bibinfo {author} {\bibfnamefont {T.}~\bibnamefont {Horiba}},\
  and\ \bibinfo {author} {\bibfnamefont {H.}~\bibnamefont {Hirai}},\ }\bibfield
   {title} {\bibinfo {title} {Calculation of core-excited and core-ionized
  states using variational quantum deflation method and applications to
  photocatalyst modeling},\ }\href@noop {} {\bibfield  {journal} {\bibinfo
  {journal} {Acs Omega}\ }\textbf {\bibinfo {volume} {7}},\ \bibinfo {pages}
  {10840--10853} (\bibinfo {year} {2022})}\BibitemShut {NoStop}%
\bibitem [{\citenamefont {Shirai}\ \emph {et~al.}(2023)\citenamefont {Shirai},
  \citenamefont {Iwakiri}, \citenamefont {Kanno}, \citenamefont {Horiba},
  \citenamefont {Omiya}, \citenamefont {Hirai},\ and\ \citenamefont
  {Koh}}]{shirai2023computational}%
  \BibitemOpen
  \bibfield  {author} {\bibinfo {author} {\bibfnamefont {S.}~\bibnamefont
  {Shirai}}, \bibinfo {author} {\bibfnamefont {H.}~\bibnamefont {Iwakiri}},
  \bibinfo {author} {\bibfnamefont {K.}~\bibnamefont {Kanno}}, \bibinfo
  {author} {\bibfnamefont {T.}~\bibnamefont {Horiba}}, \bibinfo {author}
  {\bibfnamefont {K.}~\bibnamefont {Omiya}}, \bibinfo {author} {\bibfnamefont
  {H.}~\bibnamefont {Hirai}},\ and\ \bibinfo {author} {\bibfnamefont
  {S.}~\bibnamefont {Koh}},\ }\bibfield  {title} {\bibinfo {title}
  {Computational analysis of chemical reactions using a variational quantum
  eigensolver algorithm without specifying spin multiplicity},\ }\href@noop {}
  {\bibfield  {journal} {\bibinfo  {journal} {ACS Omega}\ } (\bibinfo {year}
  {2023})}\BibitemShut {NoStop}%
\bibitem [{\citenamefont {Hirai}(2023)}]{hirai2023excited}%
  \BibitemOpen
  \bibfield  {author} {\bibinfo {author} {\bibfnamefont {H.}~\bibnamefont
  {Hirai}},\ }\bibfield  {title} {\bibinfo {title} {Excited-state molecular
  dynamics simulation based on variational quantum algorithms},\ }\href@noop {}
  {\bibfield  {journal} {\bibinfo  {journal} {Chemical Physics Letters}\
  }\textbf {\bibinfo {volume} {816}},\ \bibinfo {pages} {140404} (\bibinfo
  {year} {2023})}\BibitemShut {NoStop}%
\bibitem [{\citenamefont {Colless}\ \emph {et~al.}(2018)\citenamefont
  {Colless}, \citenamefont {Ramasesh}, \citenamefont {Dahlen}, \citenamefont
  {Blok}, \citenamefont {Kimchi-Schwartz}, \citenamefont {McClean},
  \citenamefont {Carter}, \citenamefont {de~Jong},\ and\ \citenamefont
  {Siddiqi}}]{colless2018computation}%
  \BibitemOpen
  \bibfield  {author} {\bibinfo {author} {\bibfnamefont {J.~I.}\ \bibnamefont
  {Colless}}, \bibinfo {author} {\bibfnamefont {V.~V.}\ \bibnamefont
  {Ramasesh}}, \bibinfo {author} {\bibfnamefont {D.}~\bibnamefont {Dahlen}},
  \bibinfo {author} {\bibfnamefont {M.~S.}\ \bibnamefont {Blok}}, \bibinfo
  {author} {\bibfnamefont {M.~E.}\ \bibnamefont {Kimchi-Schwartz}}, \bibinfo
  {author} {\bibfnamefont {J.~R.}\ \bibnamefont {McClean}}, \bibinfo {author}
  {\bibfnamefont {J.}~\bibnamefont {Carter}}, \bibinfo {author} {\bibfnamefont
  {W.~A.}\ \bibnamefont {de~Jong}},\ and\ \bibinfo {author} {\bibfnamefont
  {I.}~\bibnamefont {Siddiqi}},\ }\bibfield  {title} {\bibinfo {title}
  {Computation of molecular spectra on a quantum processor with an
  error-resilient algorithm},\ }\href@noop {} {\bibfield  {journal} {\bibinfo
  {journal} {Physical Review X}\ }\textbf {\bibinfo {volume} {8}},\ \bibinfo
  {pages} {011021} (\bibinfo {year} {2018})}\BibitemShut {NoStop}%
\bibitem [{\citenamefont {Kassal}\ and\ \citenamefont
  {Aspuru-Guzik}(2009)}]{kassal2009quantum}%
  \BibitemOpen
  \bibfield  {author} {\bibinfo {author} {\bibfnamefont {I.}~\bibnamefont
  {Kassal}}\ and\ \bibinfo {author} {\bibfnamefont {A.}~\bibnamefont
  {Aspuru-Guzik}},\ }\bibfield  {title} {\bibinfo {title} {Quantum algorithm
  for molecular properties and geometry optimization},\ }\href@noop {}
  {\bibfield  {journal} {\bibinfo  {journal} {The Journal of chemical physics}\
  }\textbf {\bibinfo {volume} {131}},\ \bibinfo {pages} {224102} (\bibinfo
  {year} {2009})}\BibitemShut {NoStop}%
\bibitem [{\citenamefont {Rall}(2020)}]{rall2020quantum}%
  \BibitemOpen
  \bibfield  {author} {\bibinfo {author} {\bibfnamefont {P.}~\bibnamefont
  {Rall}},\ }\bibfield  {title} {\bibinfo {title} {Quantum algorithms for
  estimating physical quantities using block encodings},\ }\href@noop {}
  {\bibfield  {journal} {\bibinfo  {journal} {Physical Review A}\ }\textbf
  {\bibinfo {volume} {102}},\ \bibinfo {pages} {022408} (\bibinfo {year}
  {2020})}\BibitemShut {NoStop}%
\bibitem [{\citenamefont {Kassal}\ \emph {et~al.}(2008)\citenamefont {Kassal},
  \citenamefont {Jordan}, \citenamefont {Love}, \citenamefont {Mohseni},\ and\
  \citenamefont {Aspuru-Guzik}}]{kassal2008polynomial}%
  \BibitemOpen
  \bibfield  {author} {\bibinfo {author} {\bibfnamefont {I.}~\bibnamefont
  {Kassal}}, \bibinfo {author} {\bibfnamefont {S.~P.}\ \bibnamefont {Jordan}},
  \bibinfo {author} {\bibfnamefont {P.~J.}\ \bibnamefont {Love}}, \bibinfo
  {author} {\bibfnamefont {M.}~\bibnamefont {Mohseni}},\ and\ \bibinfo {author}
  {\bibfnamefont {A.}~\bibnamefont {Aspuru-Guzik}},\ }\bibfield  {title}
  {\bibinfo {title} {Polynomial-time quantum algorithm for the simulation of
  chemical dynamics},\ }\href@noop {} {\bibfield  {journal} {\bibinfo
  {journal} {Proceedings of the National Academy of Sciences}\ }\textbf
  {\bibinfo {volume} {105}},\ \bibinfo {pages} {18681--18686} (\bibinfo {year}
  {2008})}\BibitemShut {NoStop}%
\bibitem [{\citenamefont {Oftelie}\ \emph {et~al.}(2020)\citenamefont
  {Oftelie}, \citenamefont {Liu}, \citenamefont {Krishnamoorthy}, \citenamefont
  {Linker}, \citenamefont {Geng}, \citenamefont {Shebib}, \citenamefont
  {Fukushima}, \citenamefont {Shimojo}, \citenamefont {Kalia}, \citenamefont
  {Nakano} \emph {et~al.}}]{oftelie2020towards}%
  \BibitemOpen
  \bibfield  {author} {\bibinfo {author} {\bibfnamefont {L.~B.}\ \bibnamefont
  {Oftelie}}, \bibinfo {author} {\bibfnamefont {K.}~\bibnamefont {Liu}},
  \bibinfo {author} {\bibfnamefont {A.}~\bibnamefont {Krishnamoorthy}},
  \bibinfo {author} {\bibfnamefont {T.}~\bibnamefont {Linker}}, \bibinfo
  {author} {\bibfnamefont {Y.}~\bibnamefont {Geng}}, \bibinfo {author}
  {\bibfnamefont {D.}~\bibnamefont {Shebib}}, \bibinfo {author} {\bibfnamefont
  {S.}~\bibnamefont {Fukushima}}, \bibinfo {author} {\bibfnamefont
  {F.}~\bibnamefont {Shimojo}}, \bibinfo {author} {\bibfnamefont {R.~K.}\
  \bibnamefont {Kalia}}, \bibinfo {author} {\bibfnamefont {A.}~\bibnamefont
  {Nakano}}, \emph {et~al.},\ }\bibfield  {title} {\bibinfo {title} {Towards
  simulation of the dynamics of materials on quantum computers},\ }\href@noop
  {} {\bibfield  {journal} {\bibinfo  {journal} {Physical Review B}\ }\textbf
  {\bibinfo {volume} {101}},\ \bibinfo {pages} {184305} (\bibinfo {year}
  {2020})}\BibitemShut {NoStop}%
\bibitem [{\citenamefont {Ollitrault}\ \emph {et~al.}(2021)\citenamefont
  {Ollitrault}, \citenamefont {Miessen},\ and\ \citenamefont
  {Tavernelli}}]{ollitrault2021molecular}%
  \BibitemOpen
  \bibfield  {author} {\bibinfo {author} {\bibfnamefont {P.~J.}\ \bibnamefont
  {Ollitrault}}, \bibinfo {author} {\bibfnamefont {A.}~\bibnamefont
  {Miessen}},\ and\ \bibinfo {author} {\bibfnamefont {I.}~\bibnamefont
  {Tavernelli}},\ }\bibfield  {title} {\bibinfo {title} {Molecular quantum
  dynamics: A quantum computing perspective},\ }\href@noop {} {\bibfield
  {journal} {\bibinfo  {journal} {Accounts of Chemical Research}\ }\textbf
  {\bibinfo {volume} {54}},\ \bibinfo {pages} {4229--4238} (\bibinfo {year}
  {2021})}\BibitemShut {NoStop}%
\bibitem [{\citenamefont {Peruzzo}\ \emph {et~al.}(2014)\citenamefont
  {Peruzzo}, \citenamefont {McClean}, \citenamefont {Shadbolt}, \citenamefont
  {Yung}, \citenamefont {Zhou}, \citenamefont {Love}, \citenamefont
  {Aspuru-Guzik},\ and\ \citenamefont {O’brien}}]{peruzzo2014variational}%
  \BibitemOpen
  \bibfield  {author} {\bibinfo {author} {\bibfnamefont {A.}~\bibnamefont
  {Peruzzo}}, \bibinfo {author} {\bibfnamefont {J.}~\bibnamefont {McClean}},
  \bibinfo {author} {\bibfnamefont {P.}~\bibnamefont {Shadbolt}}, \bibinfo
  {author} {\bibfnamefont {M.-H.}\ \bibnamefont {Yung}}, \bibinfo {author}
  {\bibfnamefont {X.-Q.}\ \bibnamefont {Zhou}}, \bibinfo {author}
  {\bibfnamefont {P.~J.}\ \bibnamefont {Love}}, \bibinfo {author}
  {\bibfnamefont {A.}~\bibnamefont {Aspuru-Guzik}},\ and\ \bibinfo {author}
  {\bibfnamefont {J.~L.}\ \bibnamefont {O’brien}},\ }\bibfield  {title}
  {\bibinfo {title} {A variational eigenvalue solver on a photonic quantum
  processor},\ }\href@noop {} {\bibfield  {journal} {\bibinfo  {journal} {Nat.
  Comm.}\ }\textbf {\bibinfo {volume} {5}},\ \bibinfo {pages} {4213} (\bibinfo
  {year} {2014})}\BibitemShut {NoStop}%
\bibitem [{\citenamefont {McClean}\ \emph {et~al.}(2016)\citenamefont
  {McClean}, \citenamefont {Romero}, \citenamefont {Babbush},\ and\
  \citenamefont {Aspuru-Guzik}}]{mcclean2016theory}%
  \BibitemOpen
  \bibfield  {author} {\bibinfo {author} {\bibfnamefont {J.~R.}\ \bibnamefont
  {McClean}}, \bibinfo {author} {\bibfnamefont {J.}~\bibnamefont {Romero}},
  \bibinfo {author} {\bibfnamefont {R.}~\bibnamefont {Babbush}},\ and\ \bibinfo
  {author} {\bibfnamefont {A.}~\bibnamefont {Aspuru-Guzik}},\ }\bibfield
  {title} {\bibinfo {title} {The theory of variational hybrid quantum-classical
  algorithms},\ }\href@noop {} {\bibfield  {journal} {\bibinfo  {journal} {New
  Journal of Physics}\ }\textbf {\bibinfo {volume} {18}},\ \bibinfo {pages}
  {023023} (\bibinfo {year} {2016})}\BibitemShut {NoStop}%
\bibitem [{\citenamefont {Cerezo}\ \emph
  {et~al.}(2021{\natexlab{a}})\citenamefont {Cerezo}, \citenamefont
  {Arrasmith}, \citenamefont {Babbush}, \citenamefont {Benjamin}, \citenamefont
  {Endo}, \citenamefont {Fujii}, \citenamefont {McClean}, \citenamefont
  {Mitarai}, \citenamefont {Yuan}, \citenamefont {Cincio} \emph
  {et~al.}}]{cerezo2021variational}%
  \BibitemOpen
  \bibfield  {author} {\bibinfo {author} {\bibfnamefont {M.}~\bibnamefont
  {Cerezo}}, \bibinfo {author} {\bibfnamefont {A.}~\bibnamefont {Arrasmith}},
  \bibinfo {author} {\bibfnamefont {R.}~\bibnamefont {Babbush}}, \bibinfo
  {author} {\bibfnamefont {S.~C.}\ \bibnamefont {Benjamin}}, \bibinfo {author}
  {\bibfnamefont {S.}~\bibnamefont {Endo}}, \bibinfo {author} {\bibfnamefont
  {K.}~\bibnamefont {Fujii}}, \bibinfo {author} {\bibfnamefont {J.~R.}\
  \bibnamefont {McClean}}, \bibinfo {author} {\bibfnamefont {K.}~\bibnamefont
  {Mitarai}}, \bibinfo {author} {\bibfnamefont {X.}~\bibnamefont {Yuan}},
  \bibinfo {author} {\bibfnamefont {L.}~\bibnamefont {Cincio}}, \emph
  {et~al.},\ }\bibfield  {title} {\bibinfo {title} {Variational quantum
  algorithms},\ }\href@noop {} {\bibfield  {journal} {\bibinfo  {journal}
  {Nature Reviews Physics}\ }\textbf {\bibinfo {volume} {3}},\ \bibinfo {pages}
  {625--644} (\bibinfo {year} {2021}{\natexlab{a}})}\BibitemShut {NoStop}%
\bibitem [{\citenamefont {Zalka}(1998)}]{zalka1998efficient}%
  \BibitemOpen
  \bibfield  {author} {\bibinfo {author} {\bibfnamefont {C.}~\bibnamefont
  {Zalka}},\ }\bibfield  {title} {\bibinfo {title} {Efficient simulation of
  quantum systems by quantum computers},\ }\href@noop {} {\bibfield  {journal}
  {\bibinfo  {journal} {Fortschritte der Physik: Progress of Physics}\ }\textbf
  {\bibinfo {volume} {46}},\ \bibinfo {pages} {877--879} (\bibinfo {year}
  {1998})}\BibitemShut {NoStop}%
\bibitem [{\citenamefont {Abrams}\ and\ \citenamefont
  {Lloyd}(1997)}]{abrams1997simulation}%
  \BibitemOpen
  \bibfield  {author} {\bibinfo {author} {\bibfnamefont {D.~S.}\ \bibnamefont
  {Abrams}}\ and\ \bibinfo {author} {\bibfnamefont {S.}~\bibnamefont {Lloyd}},\
  }\bibfield  {title} {\bibinfo {title} {Simulation of many-body fermi systems
  on a universal quantum computer},\ }\href@noop {} {\bibfield  {journal}
  {\bibinfo  {journal} {Physical Review Letters}\ }\textbf {\bibinfo {volume}
  {79}},\ \bibinfo {pages} {2586} (\bibinfo {year} {1997})}\BibitemShut
  {NoStop}%
\bibitem [{\citenamefont {Berry}\ \emph {et~al.}(2018)\citenamefont {Berry},
  \citenamefont {Kieferov{\'a}}, \citenamefont {Scherer}, \citenamefont
  {Sanders}, \citenamefont {Low}, \citenamefont {Wiebe}, \citenamefont
  {Gidney},\ and\ \citenamefont {Babbush}}]{berry2018improved}%
  \BibitemOpen
  \bibfield  {author} {\bibinfo {author} {\bibfnamefont {D.~W.}\ \bibnamefont
  {Berry}}, \bibinfo {author} {\bibfnamefont {M.}~\bibnamefont
  {Kieferov{\'a}}}, \bibinfo {author} {\bibfnamefont {A.}~\bibnamefont
  {Scherer}}, \bibinfo {author} {\bibfnamefont {Y.~R.}\ \bibnamefont
  {Sanders}}, \bibinfo {author} {\bibfnamefont {G.~H.}\ \bibnamefont {Low}},
  \bibinfo {author} {\bibfnamefont {N.}~\bibnamefont {Wiebe}}, \bibinfo
  {author} {\bibfnamefont {C.}~\bibnamefont {Gidney}},\ and\ \bibinfo {author}
  {\bibfnamefont {R.}~\bibnamefont {Babbush}},\ }\bibfield  {title} {\bibinfo
  {title} {Improved techniques for preparing eigenstates of fermionic
  hamiltonians},\ }\href@noop {} {\bibfield  {journal} {\bibinfo  {journal}
  {npj Quantum Information}\ }\textbf {\bibinfo {volume} {4}},\ \bibinfo
  {pages} {22} (\bibinfo {year} {2018})}\BibitemShut {NoStop}%
\bibitem [{\citenamefont {Babbush}\ \emph {et~al.}(2019)\citenamefont
  {Babbush}, \citenamefont {Berry}, \citenamefont {McClean},\ and\
  \citenamefont {Neven}}]{babbush2019quantum}%
  \BibitemOpen
  \bibfield  {author} {\bibinfo {author} {\bibfnamefont {R.}~\bibnamefont
  {Babbush}}, \bibinfo {author} {\bibfnamefont {D.~W.}\ \bibnamefont {Berry}},
  \bibinfo {author} {\bibfnamefont {J.~R.}\ \bibnamefont {McClean}},\ and\
  \bibinfo {author} {\bibfnamefont {H.}~\bibnamefont {Neven}},\ }\bibfield
  {title} {\bibinfo {title} {Quantum simulation of chemistry with sublinear
  scaling in basis size},\ }\href@noop {} {\bibfield  {journal} {\bibinfo
  {journal} {npj Quantum Information}\ }\textbf {\bibinfo {volume} {5}},\
  \bibinfo {pages} {92} (\bibinfo {year} {2019})}\BibitemShut {NoStop}%
\bibitem [{\citenamefont {Chan}\ \emph {et~al.}(2023)\citenamefont {Chan},
  \citenamefont {Meister}, \citenamefont {Jones}, \citenamefont {Tew},\ and\
  \citenamefont {Benjamin}}]{chan2023grid}%
  \BibitemOpen
  \bibfield  {author} {\bibinfo {author} {\bibfnamefont {H.~H.~S.}\
  \bibnamefont {Chan}}, \bibinfo {author} {\bibfnamefont {R.}~\bibnamefont
  {Meister}}, \bibinfo {author} {\bibfnamefont {T.}~\bibnamefont {Jones}},
  \bibinfo {author} {\bibfnamefont {D.~P.}\ \bibnamefont {Tew}},\ and\ \bibinfo
  {author} {\bibfnamefont {S.~C.}\ \bibnamefont {Benjamin}},\ }\bibfield
  {title} {\bibinfo {title} {Grid-based methods for chemistry simulations on a
  quantum computer},\ }\href@noop {} {\bibfield  {journal} {\bibinfo  {journal}
  {Science Advances}\ }\textbf {\bibinfo {volume} {9}},\ \bibinfo {pages}
  {eabo7484} (\bibinfo {year} {2023})}\BibitemShut {NoStop}%
\bibitem [{\citenamefont {Su}\ \emph {et~al.}(2021)\citenamefont {Su},
  \citenamefont {Berry}, \citenamefont {Wiebe}, \citenamefont {Rubin},\ and\
  \citenamefont {Babbush}}]{su2021fault}%
  \BibitemOpen
  \bibfield  {author} {\bibinfo {author} {\bibfnamefont {Y.}~\bibnamefont
  {Su}}, \bibinfo {author} {\bibfnamefont {D.~W.}\ \bibnamefont {Berry}},
  \bibinfo {author} {\bibfnamefont {N.}~\bibnamefont {Wiebe}}, \bibinfo
  {author} {\bibfnamefont {N.}~\bibnamefont {Rubin}},\ and\ \bibinfo {author}
  {\bibfnamefont {R.}~\bibnamefont {Babbush}},\ }\bibfield  {title} {\bibinfo
  {title} {Fault-tolerant quantum simulations of chemistry in first
  quantization},\ }\href@noop {} {\bibfield  {journal} {\bibinfo  {journal}
  {PRX Quantum}\ }\textbf {\bibinfo {volume} {2}},\ \bibinfo {pages} {040332}
  (\bibinfo {year} {2021})}\BibitemShut {NoStop}%
\bibitem [{\citenamefont {Hirose}\ \emph {et~al.}(2005)\citenamefont {Hirose},
  \citenamefont {Ono}, \citenamefont {Fujimoto},\ and\ \citenamefont
  {Tsukamoto}}]{hirose2005first}%
  \BibitemOpen
  \bibfield  {author} {\bibinfo {author} {\bibfnamefont {K.}~\bibnamefont
  {Hirose}}, \bibinfo {author} {\bibfnamefont {T.}~\bibnamefont {Ono}},
  \bibinfo {author} {\bibfnamefont {Y.}~\bibnamefont {Fujimoto}},\ and\
  \bibinfo {author} {\bibfnamefont {S.}~\bibnamefont {Tsukamoto}},\ }\href@noop
  {} {\emph {\bibinfo {title} {First-principles calculations in real-space
  formalism: electronic configurations and transport properties of
  nanostructures}}}\ (\bibinfo  {publisher} {World Scientific},\ \bibinfo
  {year} {2005})\BibitemShut {NoStop}%
\bibitem [{\citenamefont {Ohba}\ \emph {et~al.}(2012)\citenamefont {Ohba},
  \citenamefont {Ogata}, \citenamefont {Kouno}, \citenamefont {Tamura},\ and\
  \citenamefont {Kobayashi}}]{ohba2012linear}%
  \BibitemOpen
  \bibfield  {author} {\bibinfo {author} {\bibfnamefont {N.}~\bibnamefont
  {Ohba}}, \bibinfo {author} {\bibfnamefont {S.}~\bibnamefont {Ogata}},
  \bibinfo {author} {\bibfnamefont {T.}~\bibnamefont {Kouno}}, \bibinfo
  {author} {\bibfnamefont {T.}~\bibnamefont {Tamura}},\ and\ \bibinfo {author}
  {\bibfnamefont {R.}~\bibnamefont {Kobayashi}},\ }\bibfield  {title} {\bibinfo
  {title} {Linear scaling algorithm of real-space density functional theory of
  electrons with correlated overlapping domains},\ }\href@noop {} {\bibfield
  {journal} {\bibinfo  {journal} {Computer Physics Communications}\ }\textbf
  {\bibinfo {volume} {183}},\ \bibinfo {pages} {1664--1673} (\bibinfo {year}
  {2012})}\BibitemShut {NoStop}%
\bibitem [{\citenamefont {Childs}\ \emph {et~al.}(2022)\citenamefont {Childs},
  \citenamefont {Leng}, \citenamefont {Li}, \citenamefont {Liu},\ and\
  \citenamefont {Zhang}}]{childs2022quantum}%
  \BibitemOpen
  \bibfield  {author} {\bibinfo {author} {\bibfnamefont {A.~M.}\ \bibnamefont
  {Childs}}, \bibinfo {author} {\bibfnamefont {J.}~\bibnamefont {Leng}},
  \bibinfo {author} {\bibfnamefont {T.}~\bibnamefont {Li}}, \bibinfo {author}
  {\bibfnamefont {J.-P.}\ \bibnamefont {Liu}},\ and\ \bibinfo {author}
  {\bibfnamefont {C.}~\bibnamefont {Zhang}},\ }\bibfield  {title} {\bibinfo
  {title} {Quantum simulation of real-space dynamics},\ }\href@noop {}
  {\bibfield  {journal} {\bibinfo  {journal} {Quantum}\ }\textbf {\bibinfo
  {volume} {6}},\ \bibinfo {pages} {860} (\bibinfo {year} {2022})}\BibitemShut
  {NoStop}%
\bibitem [{\citenamefont {Nielsen}\ and\ \citenamefont
  {Chuang}(2002)}]{nielsen2002quantum}%
  \BibitemOpen
  \bibfield  {author} {\bibinfo {author} {\bibfnamefont {M.~A.}\ \bibnamefont
  {Nielsen}}\ and\ \bibinfo {author} {\bibfnamefont {I.}~\bibnamefont
  {Chuang}},\ }\href@noop {} {\bibinfo {title} {Quantum computation and quantum
  information}} (\bibinfo {year} {2002})\BibitemShut {NoStop}%
\bibitem [{\citenamefont {Holmes}\ and\ \citenamefont
  {Matsuura}(2020)}]{holmes2020efficient}%
  \BibitemOpen
  \bibfield  {author} {\bibinfo {author} {\bibfnamefont {A.}~\bibnamefont
  {Holmes}}\ and\ \bibinfo {author} {\bibfnamefont {A.~Y.}\ \bibnamefont
  {Matsuura}},\ }\bibfield  {title} {\bibinfo {title} {Efficient quantum
  circuits for accurate state preparation of smooth, differentiable
  functions},\ }in\ \href@noop {} {\emph {\bibinfo {booktitle} {2020 IEEE
  International Conference on Quantum Computing and Engineering (QCE)}}}\
  (\bibinfo {organization} {IEEE},\ \bibinfo {year} {2020})\ pp.\ \bibinfo
  {pages} {169--179}\BibitemShut {NoStop}%
\bibitem [{\citenamefont {Koppe}\ and\ \citenamefont
  {Wolf}(2023)}]{koppe2023amplitude}%
  \BibitemOpen
  \bibfield  {author} {\bibinfo {author} {\bibfnamefont {J.}~\bibnamefont
  {Koppe}}\ and\ \bibinfo {author} {\bibfnamefont {M.-O.}\ \bibnamefont
  {Wolf}},\ }\bibfield  {title} {\bibinfo {title} {Amplitude-based
  implementation of the unit step function on a quantum computer},\ }\href@noop
  {} {\bibfield  {journal} {\bibinfo  {journal} {Physical Review A}\ }\textbf
  {\bibinfo {volume} {107}},\ \bibinfo {pages} {022606} (\bibinfo {year}
  {2023})}\BibitemShut {NoStop}%
\bibitem [{\citenamefont {Ollitrault}\ \emph {et~al.}(2020)\citenamefont
  {Ollitrault}, \citenamefont {Mazzola},\ and\ \citenamefont
  {Tavernelli}}]{ollitrault2020nonadiabatic}%
  \BibitemOpen
  \bibfield  {author} {\bibinfo {author} {\bibfnamefont {P.~J.}\ \bibnamefont
  {Ollitrault}}, \bibinfo {author} {\bibfnamefont {G.}~\bibnamefont
  {Mazzola}},\ and\ \bibinfo {author} {\bibfnamefont {I.}~\bibnamefont
  {Tavernelli}},\ }\bibfield  {title} {\bibinfo {title} {Nonadiabatic molecular
  quantum dynamics with quantum computers},\ }\href@noop {} {\bibfield
  {journal} {\bibinfo  {journal} {Physical Review Letters}\ }\textbf {\bibinfo
  {volume} {125}},\ \bibinfo {pages} {260511} (\bibinfo {year}
  {2020})}\BibitemShut {NoStop}%
\bibitem [{\citenamefont {Nakaji}\ \emph {et~al.}(2022)\citenamefont {Nakaji},
  \citenamefont {Uno}, \citenamefont {Suzuki}, \citenamefont {Raymond},
  \citenamefont {Onodera}, \citenamefont {Tanaka}, \citenamefont {Tezuka},
  \citenamefont {Mitsuda},\ and\ \citenamefont
  {Yamamoto}}]{nakaji2022approximate}%
  \BibitemOpen
  \bibfield  {author} {\bibinfo {author} {\bibfnamefont {K.}~\bibnamefont
  {Nakaji}}, \bibinfo {author} {\bibfnamefont {S.}~\bibnamefont {Uno}},
  \bibinfo {author} {\bibfnamefont {Y.}~\bibnamefont {Suzuki}}, \bibinfo
  {author} {\bibfnamefont {R.}~\bibnamefont {Raymond}}, \bibinfo {author}
  {\bibfnamefont {T.}~\bibnamefont {Onodera}}, \bibinfo {author} {\bibfnamefont
  {T.}~\bibnamefont {Tanaka}}, \bibinfo {author} {\bibfnamefont
  {H.}~\bibnamefont {Tezuka}}, \bibinfo {author} {\bibfnamefont
  {N.}~\bibnamefont {Mitsuda}},\ and\ \bibinfo {author} {\bibfnamefont
  {N.}~\bibnamefont {Yamamoto}},\ }\bibfield  {title} {\bibinfo {title}
  {Approximate amplitude encoding in shallow parameterized quantum circuits and
  its application to financial market indicators},\ }\href@noop {} {\bibfield
  {journal} {\bibinfo  {journal} {Physical Review Research}\ }\textbf {\bibinfo
  {volume} {4}},\ \bibinfo {pages} {023136} (\bibinfo {year}
  {2022})}\BibitemShut {NoStop}%
\bibitem [{\citenamefont {Halder}\ \emph {et~al.}(2023)\citenamefont {Halder},
  \citenamefont {Prasannaa}, \citenamefont {Agarawal},\ and\ \citenamefont
  {Maitra}}]{halder2023iterative}%
  \BibitemOpen
  \bibfield  {author} {\bibinfo {author} {\bibfnamefont {D.}~\bibnamefont
  {Halder}}, \bibinfo {author} {\bibfnamefont {V.~S.}\ \bibnamefont
  {Prasannaa}}, \bibinfo {author} {\bibfnamefont {V.}~\bibnamefont
  {Agarawal}},\ and\ \bibinfo {author} {\bibfnamefont {R.}~\bibnamefont
  {Maitra}},\ }\bibfield  {title} {\bibinfo {title} {Iterative quantum phase
  estimation with variationally prepared reference state},\ }\href@noop {}
  {\bibfield  {journal} {\bibinfo  {journal} {International Journal of Quantum
  Chemistry}\ ,\ \bibinfo {pages} {e27021}} (\bibinfo {year}
  {2023})}\BibitemShut {NoStop}%
\bibitem [{\citenamefont {Dong}\ \emph {et~al.}(2022)\citenamefont {Dong},
  \citenamefont {Lin},\ and\ \citenamefont {Tong}}]{dong2022ground}%
  \BibitemOpen
  \bibfield  {author} {\bibinfo {author} {\bibfnamefont {Y.}~\bibnamefont
  {Dong}}, \bibinfo {author} {\bibfnamefont {L.}~\bibnamefont {Lin}},\ and\
  \bibinfo {author} {\bibfnamefont {Y.}~\bibnamefont {Tong}},\ }\bibfield
  {title} {\bibinfo {title} {Ground-state preparation and energy estimation on
  early fault-tolerant quantum computers via quantum eigenvalue transformation
  of unitary matrices},\ }\href@noop {} {\bibfield  {journal} {\bibinfo
  {journal} {PRX Quantum}\ }\textbf {\bibinfo {volume} {3}},\ \bibinfo {pages}
  {040305} (\bibinfo {year} {2022})}\BibitemShut {NoStop}%
\bibitem [{\citenamefont {Durstenfeld}(1964)}]{durstenfeld1964algorithm}%
  \BibitemOpen
  \bibfield  {author} {\bibinfo {author} {\bibfnamefont {R.}~\bibnamefont
  {Durstenfeld}},\ }\bibfield  {title} {\bibinfo {title} {Algorithm 235: random
  permutation},\ }\href@noop {} {\bibfield  {journal} {\bibinfo  {journal}
  {Communications of the ACM}\ }\textbf {\bibinfo {volume} {7}},\ \bibinfo
  {pages} {420} (\bibinfo {year} {1964})}\BibitemShut {NoStop}%
\bibitem [{\citenamefont {Jones}(2003)}]{jones2003robust}%
  \BibitemOpen
  \bibfield  {author} {\bibinfo {author} {\bibfnamefont {J.~A.}\ \bibnamefont
  {Jones}},\ }\bibfield  {title} {\bibinfo {title} {Robust ising gates for
  practical quantum computation},\ }\href@noop {} {\bibfield  {journal}
  {\bibinfo  {journal} {Physical Review A}\ }\textbf {\bibinfo {volume} {67}},\
  \bibinfo {pages} {012317} (\bibinfo {year} {2003})}\BibitemShut {NoStop}%
\bibitem [{\citenamefont {Whitfield}\ \emph {et~al.}(2011)\citenamefont
  {Whitfield}, \citenamefont {Biamonte},\ and\ \citenamefont
  {Aspuru-Guzik}}]{whitfield2011simulation}%
  \BibitemOpen
  \bibfield  {author} {\bibinfo {author} {\bibfnamefont {J.~D.}\ \bibnamefont
  {Whitfield}}, \bibinfo {author} {\bibfnamefont {J.}~\bibnamefont
  {Biamonte}},\ and\ \bibinfo {author} {\bibfnamefont {A.}~\bibnamefont
  {Aspuru-Guzik}},\ }\bibfield  {title} {\bibinfo {title} {Simulation of
  electronic structure hamiltonians using quantum computers},\ }\href@noop {}
  {\bibfield  {journal} {\bibinfo  {journal} {Molecular Physics}\ }\textbf
  {\bibinfo {volume} {109}},\ \bibinfo {pages} {735--750} (\bibinfo {year}
  {2011})}\BibitemShut {NoStop}%
\bibitem [{\citenamefont {K\"{u}hn}\ \emph {et~al.}(2019)\citenamefont
  {K\"{u}hn}, \citenamefont {Zanker}, \citenamefont {Deglmann}, \citenamefont
  {Marthaler},\ and\ \citenamefont {Wei{\ss}}}]{kuhn2019accuracy}%
  \BibitemOpen
  \bibfield  {author} {\bibinfo {author} {\bibfnamefont {M.}~\bibnamefont
  {K\"{u}hn}}, \bibinfo {author} {\bibfnamefont {S.}~\bibnamefont {Zanker}},
  \bibinfo {author} {\bibfnamefont {P.}~\bibnamefont {Deglmann}}, \bibinfo
  {author} {\bibfnamefont {M.}~\bibnamefont {Marthaler}},\ and\ \bibinfo
  {author} {\bibfnamefont {H.}~\bibnamefont {Wei{\ss}}},\ }\bibfield  {title}
  {\bibinfo {title} {Accuracy and resource estimations for quantum chemistry on
  a near-term quantum computer},\ }\href@noop {} {\bibfield  {journal}
  {\bibinfo  {journal} {Journal of chemical theory and computation}\ }\textbf
  {\bibinfo {volume} {15}},\ \bibinfo {pages} {4764--4780} (\bibinfo {year}
  {2019})}\BibitemShut {NoStop}%
\bibitem [{\citenamefont {Ibe}\ \emph {et~al.}(2022)\citenamefont {Ibe},
  \citenamefont {Nakagawa}, \citenamefont {Earnest}, \citenamefont {Yamamoto},
  \citenamefont {Mitarai}, \citenamefont {Gao},\ and\ \citenamefont
  {Kobayashi}}]{ibe2022calculating}%
  \BibitemOpen
  \bibfield  {author} {\bibinfo {author} {\bibfnamefont {Y.}~\bibnamefont
  {Ibe}}, \bibinfo {author} {\bibfnamefont {Y.~O.}\ \bibnamefont {Nakagawa}},
  \bibinfo {author} {\bibfnamefont {N.}~\bibnamefont {Earnest}}, \bibinfo
  {author} {\bibfnamefont {T.}~\bibnamefont {Yamamoto}}, \bibinfo {author}
  {\bibfnamefont {K.}~\bibnamefont {Mitarai}}, \bibinfo {author} {\bibfnamefont
  {Q.}~\bibnamefont {Gao}},\ and\ \bibinfo {author} {\bibfnamefont
  {T.}~\bibnamefont {Kobayashi}},\ }\bibfield  {title} {\bibinfo {title}
  {Calculating transition amplitudes by variational quantum deflation},\
  }\href@noop {} {\bibfield  {journal} {\bibinfo  {journal} {Physical Review
  Research}\ }\textbf {\bibinfo {volume} {4}},\ \bibinfo {pages} {013173}
  (\bibinfo {year} {2022})}\BibitemShut {NoStop}%
\bibitem [{\citenamefont {Kandala}\ \emph {et~al.}(2017)\citenamefont
  {Kandala}, \citenamefont {Mezzacapo}, \citenamefont {Temme}, \citenamefont
  {Takita}, \citenamefont {Brink}, \citenamefont {Chow},\ and\ \citenamefont
  {Gambetta}}]{kandala2017hardware}%
  \BibitemOpen
  \bibfield  {author} {\bibinfo {author} {\bibfnamefont {A.}~\bibnamefont
  {Kandala}}, \bibinfo {author} {\bibfnamefont {A.}~\bibnamefont {Mezzacapo}},
  \bibinfo {author} {\bibfnamefont {K.}~\bibnamefont {Temme}}, \bibinfo
  {author} {\bibfnamefont {M.}~\bibnamefont {Takita}}, \bibinfo {author}
  {\bibfnamefont {M.}~\bibnamefont {Brink}}, \bibinfo {author} {\bibfnamefont
  {J.~M.}\ \bibnamefont {Chow}},\ and\ \bibinfo {author} {\bibfnamefont
  {J.~M.}\ \bibnamefont {Gambetta}},\ }\bibfield  {title} {\bibinfo {title}
  {Hardware-efficient variational quantum eigensolver for small molecules and
  quantum magnets},\ }\href@noop {} {\bibfield  {journal} {\bibinfo  {journal}
  {Nature}\ }\textbf {\bibinfo {volume} {549}},\ \bibinfo {pages} {242--246}
  (\bibinfo {year} {2017})}\BibitemShut {NoStop}%
\bibitem [{\citenamefont {Shee}\ \emph {et~al.}(2022)\citenamefont {Shee},
  \citenamefont {Tsai}, \citenamefont {Hong}, \citenamefont {Cheng},\ and\
  \citenamefont {Goan}}]{shee2022qubit}%
  \BibitemOpen
  \bibfield  {author} {\bibinfo {author} {\bibfnamefont {Y.}~\bibnamefont
  {Shee}}, \bibinfo {author} {\bibfnamefont {P.-K.}\ \bibnamefont {Tsai}},
  \bibinfo {author} {\bibfnamefont {C.-L.}\ \bibnamefont {Hong}}, \bibinfo
  {author} {\bibfnamefont {H.-C.}\ \bibnamefont {Cheng}},\ and\ \bibinfo
  {author} {\bibfnamefont {H.-S.}\ \bibnamefont {Goan}},\ }\bibfield  {title}
  {\bibinfo {title} {Qubit-efficient encoding scheme for quantum simulations of
  electronic structure},\ }\href@noop {} {\bibfield  {journal} {\bibinfo
  {journal} {Physical Review Research}\ }\textbf {\bibinfo {volume} {4}},\
  \bibinfo {pages} {023154} (\bibinfo {year} {2022})}\BibitemShut {NoStop}%
\bibitem [{\citenamefont {Kingma}\ and\ \citenamefont
  {Ba}(2014)}]{kingma2014adam}%
  \BibitemOpen
  \bibfield  {author} {\bibinfo {author} {\bibfnamefont {D.~P.}\ \bibnamefont
  {Kingma}}\ and\ \bibinfo {author} {\bibfnamefont {J.}~\bibnamefont {Ba}},\
  }\bibfield  {title} {\bibinfo {title} {Adam: A method for stochastic
  optimization},\ }\href@noop {} {\bibfield  {journal} {\bibinfo  {journal}
  {arXiv preprint arXiv:1412.6980}\ } (\bibinfo {year} {2014})}\BibitemShut
  {NoStop}%
\bibitem [{\citenamefont {Jensen}(2017)}]{jensen2017introduction}%
  \BibitemOpen
  \bibfield  {author} {\bibinfo {author} {\bibfnamefont {F.}~\bibnamefont
  {Jensen}},\ }\href@noop {} {\emph {\bibinfo {title} {Introduction to
  computational chemistry}}}\ (\bibinfo  {publisher} {John wiley \& sons},\
  \bibinfo {year} {2017})\BibitemShut {NoStop}%
\bibitem [{\citenamefont {Helgaker}\ \emph {et~al.}(2013)\citenamefont
  {Helgaker}, \citenamefont {Jorgensen},\ and\ \citenamefont
  {Olsen}}]{helgaker2013molecular}%
  \BibitemOpen
  \bibfield  {author} {\bibinfo {author} {\bibfnamefont {T.}~\bibnamefont
  {Helgaker}}, \bibinfo {author} {\bibfnamefont {P.}~\bibnamefont
  {Jorgensen}},\ and\ \bibinfo {author} {\bibfnamefont {J.}~\bibnamefont
  {Olsen}},\ }\href@noop {} {\emph {\bibinfo {title} {Molecular
  electronic-structure theory}}}\ (\bibinfo  {publisher} {John Wiley \& Sons},\
  \bibinfo {year} {2013})\BibitemShut {NoStop}%
\bibitem [{\citenamefont {Hirai}\ \emph {et~al.}(2022)\citenamefont {Hirai},
  \citenamefont {Horiba}, \citenamefont {Shirai}, \citenamefont {Kanno},
  \citenamefont {Omiya}, \citenamefont {Nakagawa},\ and\ \citenamefont
  {Koh}}]{hirai2022molecular}%
  \BibitemOpen
  \bibfield  {author} {\bibinfo {author} {\bibfnamefont {H.}~\bibnamefont
  {Hirai}}, \bibinfo {author} {\bibfnamefont {T.}~\bibnamefont {Horiba}},
  \bibinfo {author} {\bibfnamefont {S.}~\bibnamefont {Shirai}}, \bibinfo
  {author} {\bibfnamefont {K.}~\bibnamefont {Kanno}}, \bibinfo {author}
  {\bibfnamefont {K.}~\bibnamefont {Omiya}}, \bibinfo {author} {\bibfnamefont
  {Y.~O.}\ \bibnamefont {Nakagawa}},\ and\ \bibinfo {author} {\bibfnamefont
  {S.}~\bibnamefont {Koh}},\ }\bibfield  {title} {\bibinfo {title} {Molecular
  structure optimization based on electrons--nuclei quantum dynamics
  computation},\ }\href@noop {} {\bibfield  {journal} {\bibinfo  {journal} {ACS
  omega}\ }\textbf {\bibinfo {volume} {7}},\ \bibinfo {pages} {19784--19793}
  (\bibinfo {year} {2022})}\BibitemShut {NoStop}%
\bibitem [{\citenamefont {Molina-Esp{\'\i}ritu}\ \emph
  {et~al.}(2015)\citenamefont {Molina-Esp{\'\i}ritu}, \citenamefont {Esquivel},
  \citenamefont {L{\'o}pez-Rosa},\ and\ \citenamefont
  {Dehesa}}]{molina2015quantum}%
  \BibitemOpen
  \bibfield  {author} {\bibinfo {author} {\bibfnamefont {M.}~\bibnamefont
  {Molina-Esp{\'\i}ritu}}, \bibinfo {author} {\bibfnamefont {R.}~\bibnamefont
  {Esquivel}}, \bibinfo {author} {\bibfnamefont {S.}~\bibnamefont
  {L{\'o}pez-Rosa}},\ and\ \bibinfo {author} {\bibfnamefont {J.}~\bibnamefont
  {Dehesa}},\ }\bibfield  {title} {\bibinfo {title} {Quantum entanglement and
  chemical reactivity},\ }\href@noop {} {\bibfield  {journal} {\bibinfo
  {journal} {Journal of Chemical Theory and Computation}\ }\textbf {\bibinfo
  {volume} {11}},\ \bibinfo {pages} {5144--5151} (\bibinfo {year}
  {2015})}\BibitemShut {NoStop}%
\bibitem [{\citenamefont {Brandejs}\ \emph {et~al.}(2019)\citenamefont
  {Brandejs}, \citenamefont {Veis}, \citenamefont {Szalay}, \citenamefont
  {Barcza}, \citenamefont {Pittner},\ and\ \citenamefont
  {Legeza}}]{brandejs2019quantum}%
  \BibitemOpen
  \bibfield  {author} {\bibinfo {author} {\bibfnamefont {J.}~\bibnamefont
  {Brandejs}}, \bibinfo {author} {\bibfnamefont {L.}~\bibnamefont {Veis}},
  \bibinfo {author} {\bibfnamefont {S.}~\bibnamefont {Szalay}}, \bibinfo
  {author} {\bibfnamefont {G.}~\bibnamefont {Barcza}}, \bibinfo {author}
  {\bibfnamefont {J.}~\bibnamefont {Pittner}},\ and\ \bibinfo {author}
  {\bibfnamefont {{\"O}.}~\bibnamefont {Legeza}},\ }\bibfield  {title}
  {\bibinfo {title} {Quantum information-based analysis of electron-deficient
  bonds},\ }\href@noop {} {\bibfield  {journal} {\bibinfo  {journal} {The
  Journal of Chemical Physics}\ }\textbf {\bibinfo {volume} {150}},\ \bibinfo
  {pages} {204117} (\bibinfo {year} {2019})}\BibitemShut {NoStop}%
\bibitem [{\citenamefont {Esquivel}\ \emph {et~al.}(2011)\citenamefont
  {Esquivel}, \citenamefont {Flores-Gallegos}, \citenamefont
  {Molina-Esp{\'\i}ritu}, \citenamefont {Plastino}, \citenamefont {Angulo},
  \citenamefont {Antol{\'\i}n},\ and\ \citenamefont
  {Dehesa}}]{esquivel2011quantum}%
  \BibitemOpen
  \bibfield  {author} {\bibinfo {author} {\bibfnamefont {R.~O.}\ \bibnamefont
  {Esquivel}}, \bibinfo {author} {\bibfnamefont {N.}~\bibnamefont
  {Flores-Gallegos}}, \bibinfo {author} {\bibfnamefont {M.}~\bibnamefont
  {Molina-Esp{\'\i}ritu}}, \bibinfo {author} {\bibfnamefont {A.}~\bibnamefont
  {Plastino}}, \bibinfo {author} {\bibfnamefont {J.~C.}\ \bibnamefont
  {Angulo}}, \bibinfo {author} {\bibfnamefont {J.}~\bibnamefont
  {Antol{\'\i}n}},\ and\ \bibinfo {author} {\bibfnamefont {J.~S.}\ \bibnamefont
  {Dehesa}},\ }\bibfield  {title} {\bibinfo {title} {Quantum entanglement and
  the dissociation process of diatomic molecules},\ }\href@noop {} {\bibfield
  {journal} {\bibinfo  {journal} {Journal of Physics B: Atomic, Molecular and
  Optical Physics}\ }\textbf {\bibinfo {volume} {44}},\ \bibinfo {pages}
  {175101} (\bibinfo {year} {2011})}\BibitemShut {NoStop}%
\bibitem [{\citenamefont {Rissler}\ \emph {et~al.}(2006)\citenamefont
  {Rissler}, \citenamefont {Noack},\ and\ \citenamefont
  {White}}]{rissler2006measuring}%
  \BibitemOpen
  \bibfield  {author} {\bibinfo {author} {\bibfnamefont {J.}~\bibnamefont
  {Rissler}}, \bibinfo {author} {\bibfnamefont {R.~M.}\ \bibnamefont {Noack}},\
  and\ \bibinfo {author} {\bibfnamefont {S.~R.}\ \bibnamefont {White}},\
  }\bibfield  {title} {\bibinfo {title} {Measuring orbital interaction using
  quantum information theory},\ }\href@noop {} {\bibfield  {journal} {\bibinfo
  {journal} {Chemical Physics}\ }\textbf {\bibinfo {volume} {323}},\ \bibinfo
  {pages} {519--531} (\bibinfo {year} {2006})}\BibitemShut {NoStop}%
\bibitem [{\citenamefont {Boguslawski}\ \emph {et~al.}(2013)\citenamefont
  {Boguslawski}, \citenamefont {Tecmer}, \citenamefont {Barcza}, \citenamefont
  {Legeza},\ and\ \citenamefont {Reiher}}]{boguslawski2013orbital}%
  \BibitemOpen
  \bibfield  {author} {\bibinfo {author} {\bibfnamefont {K.}~\bibnamefont
  {Boguslawski}}, \bibinfo {author} {\bibfnamefont {P.}~\bibnamefont {Tecmer}},
  \bibinfo {author} {\bibfnamefont {G.}~\bibnamefont {Barcza}}, \bibinfo
  {author} {\bibfnamefont {O.}~\bibnamefont {Legeza}},\ and\ \bibinfo {author}
  {\bibfnamefont {M.}~\bibnamefont {Reiher}},\ }\bibfield  {title} {\bibinfo
  {title} {Orbital entanglement in bond-formation processes},\ }\href@noop {}
  {\bibfield  {journal} {\bibinfo  {journal} {Journal of Chemical Theory and
  Computation}\ }\textbf {\bibinfo {volume} {9}},\ \bibinfo {pages}
  {2959--2973} (\bibinfo {year} {2013})}\BibitemShut {NoStop}%
\bibitem [{\citenamefont {Lin}\ \emph {et~al.}(2013)\citenamefont {Lin},
  \citenamefont {Lin},\ and\ \citenamefont {Ho}}]{lin2013spatial}%
  \BibitemOpen
  \bibfield  {author} {\bibinfo {author} {\bibfnamefont {Y.-C.}\ \bibnamefont
  {Lin}}, \bibinfo {author} {\bibfnamefont {C.-Y.}\ \bibnamefont {Lin}},\ and\
  \bibinfo {author} {\bibfnamefont {Y.~K.}\ \bibnamefont {Ho}},\ }\bibfield
  {title} {\bibinfo {title} {Spatial entanglement in two-electron atomic
  systems},\ }\href@noop {} {\bibfield  {journal} {\bibinfo  {journal}
  {Physical Review A}\ }\textbf {\bibinfo {volume} {87}},\ \bibinfo {pages}
  {022316} (\bibinfo {year} {2013})}\BibitemShut {NoStop}%
\bibitem [{\citenamefont {Soh}\ \emph {et~al.}(2019)\citenamefont {Soh},
  \citenamefont {Klotz}, \citenamefont {Robertson-Anderson},\ and\
  \citenamefont {Doyle}}]{soh2019long}%
  \BibitemOpen
  \bibfield  {author} {\bibinfo {author} {\bibfnamefont {B.~W.}\ \bibnamefont
  {Soh}}, \bibinfo {author} {\bibfnamefont {A.~R.}\ \bibnamefont {Klotz}},
  \bibinfo {author} {\bibfnamefont {R.~M.}\ \bibnamefont
  {Robertson-Anderson}},\ and\ \bibinfo {author} {\bibfnamefont {P.~S.}\
  \bibnamefont {Doyle}},\ }\bibfield  {title} {\bibinfo {title} {Long-lived
  self-entanglements in ring polymers},\ }\href@noop {} {\bibfield  {journal}
  {\bibinfo  {journal} {Physical review letters}\ }\textbf {\bibinfo {volume}
  {123}},\ \bibinfo {pages} {048002} (\bibinfo {year} {2019})}\BibitemShut
  {NoStop}%
\bibitem [{\citenamefont {Zhu}\ \emph {et~al.}(2020)\citenamefont {Zhu},
  \citenamefont {Huang}, \citenamefont {He},\ and\ \citenamefont
  {Wen}}]{zhu2020entanglement}%
  \BibitemOpen
  \bibfield  {author} {\bibinfo {author} {\bibfnamefont {W.}~\bibnamefont
  {Zhu}}, \bibinfo {author} {\bibfnamefont {Z.}~\bibnamefont {Huang}}, \bibinfo
  {author} {\bibfnamefont {Y.-C.}\ \bibnamefont {He}},\ and\ \bibinfo {author}
  {\bibfnamefont {X.}~\bibnamefont {Wen}},\ }\bibfield  {title} {\bibinfo
  {title} {Entanglement hamiltonian of many-body dynamics in strongly
  correlated systems},\ }\href@noop {} {\bibfield  {journal} {\bibinfo
  {journal} {Physical Review Letters}\ }\textbf {\bibinfo {volume} {124}},\
  \bibinfo {pages} {100605} (\bibinfo {year} {2020})}\BibitemShut {NoStop}%
\bibitem [{\citenamefont {Lanat{\`a}}\ \emph {et~al.}(2014)\citenamefont
  {Lanat{\`a}}, \citenamefont {Strand}, \citenamefont {Yao},\ and\
  \citenamefont {Kotliar}}]{lanata2014principle}%
  \BibitemOpen
  \bibfield  {author} {\bibinfo {author} {\bibfnamefont {N.}~\bibnamefont
  {Lanat{\`a}}}, \bibinfo {author} {\bibfnamefont {H.~U.}\ \bibnamefont
  {Strand}}, \bibinfo {author} {\bibfnamefont {Y.}~\bibnamefont {Yao}},\ and\
  \bibinfo {author} {\bibfnamefont {G.}~\bibnamefont {Kotliar}},\ }\bibfield
  {title} {\bibinfo {title} {Principle of maximum entanglement entropy and
  local physics of strongly correlated materials},\ }\href@noop {} {\bibfield
  {journal} {\bibinfo  {journal} {Physical review letters}\ }\textbf {\bibinfo
  {volume} {113}},\ \bibinfo {pages} {036402} (\bibinfo {year}
  {2014})}\BibitemShut {NoStop}%
\bibitem [{\citenamefont {Chen}\ \emph {et~al.}(2019)\citenamefont {Chen},
  \citenamefont {Hashimoto}, \citenamefont {He}, \citenamefont {Song},
  \citenamefont {Xu}, \citenamefont {He}, \citenamefont {Devereaux},
  \citenamefont {Eisaki}, \citenamefont {Lu}, \citenamefont {Zaanen} \emph
  {et~al.}}]{chen2019incoherent}%
  \BibitemOpen
  \bibfield  {author} {\bibinfo {author} {\bibfnamefont {S.-D.}\ \bibnamefont
  {Chen}}, \bibinfo {author} {\bibfnamefont {M.}~\bibnamefont {Hashimoto}},
  \bibinfo {author} {\bibfnamefont {Y.}~\bibnamefont {He}}, \bibinfo {author}
  {\bibfnamefont {D.}~\bibnamefont {Song}}, \bibinfo {author} {\bibfnamefont
  {K.-J.}\ \bibnamefont {Xu}}, \bibinfo {author} {\bibfnamefont {J.-F.}\
  \bibnamefont {He}}, \bibinfo {author} {\bibfnamefont {T.~P.}\ \bibnamefont
  {Devereaux}}, \bibinfo {author} {\bibfnamefont {H.}~\bibnamefont {Eisaki}},
  \bibinfo {author} {\bibfnamefont {D.-H.}\ \bibnamefont {Lu}}, \bibinfo
  {author} {\bibfnamefont {J.}~\bibnamefont {Zaanen}}, \emph {et~al.},\
  }\bibfield  {title} {\bibinfo {title} {Incoherent strange metal sharply
  bounded by a critical doping in bi2212},\ }\href@noop {} {\bibfield
  {journal} {\bibinfo  {journal} {Science}\ }\textbf {\bibinfo {volume}
  {366}},\ \bibinfo {pages} {1099--1102} (\bibinfo {year} {2019})}\BibitemShut
  {NoStop}%
\bibitem [{\citenamefont {Rebentrost}\ \emph {et~al.}(2018)\citenamefont
  {Rebentrost}, \citenamefont {Steffens}, \citenamefont {Marvian},\ and\
  \citenamefont {Lloyd}}]{rebentrost2018quantum}%
  \BibitemOpen
  \bibfield  {author} {\bibinfo {author} {\bibfnamefont {P.}~\bibnamefont
  {Rebentrost}}, \bibinfo {author} {\bibfnamefont {A.}~\bibnamefont
  {Steffens}}, \bibinfo {author} {\bibfnamefont {I.}~\bibnamefont {Marvian}},\
  and\ \bibinfo {author} {\bibfnamefont {S.}~\bibnamefont {Lloyd}},\ }\bibfield
   {title} {\bibinfo {title} {Quantum singular-value decomposition of nonsparse
  low-rank matrices},\ }\href@noop {} {\bibfield  {journal} {\bibinfo
  {journal} {Physical review A}\ }\textbf {\bibinfo {volume} {97}},\ \bibinfo
  {pages} {012327} (\bibinfo {year} {2018})}\BibitemShut {NoStop}%
\bibitem [{\citenamefont {Bravo-Prieto}\ \emph {et~al.}(2020)\citenamefont
  {Bravo-Prieto}, \citenamefont {Garc{\'\i}a-Mart{\'\i}n},\ and\ \citenamefont
  {Latorre}}]{bravo2020quantum}%
  \BibitemOpen
  \bibfield  {author} {\bibinfo {author} {\bibfnamefont {C.}~\bibnamefont
  {Bravo-Prieto}}, \bibinfo {author} {\bibfnamefont {D.}~\bibnamefont
  {Garc{\'\i}a-Mart{\'\i}n}},\ and\ \bibinfo {author} {\bibfnamefont {J.~I.}\
  \bibnamefont {Latorre}},\ }\bibfield  {title} {\bibinfo {title} {Quantum
  singular value decomposer},\ }\href@noop {} {\bibfield  {journal} {\bibinfo
  {journal} {Physical Review A}\ }\textbf {\bibinfo {volume} {101}},\ \bibinfo
  {pages} {062310} (\bibinfo {year} {2020})}\BibitemShut {NoStop}%
\bibitem [{\citenamefont {Wang}\ \emph
  {et~al.}(2021{\natexlab{a}})\citenamefont {Wang}, \citenamefont {Song},\ and\
  \citenamefont {Wang}}]{wang2021variational}%
  \BibitemOpen
  \bibfield  {author} {\bibinfo {author} {\bibfnamefont {X.}~\bibnamefont
  {Wang}}, \bibinfo {author} {\bibfnamefont {Z.}~\bibnamefont {Song}},\ and\
  \bibinfo {author} {\bibfnamefont {Y.}~\bibnamefont {Wang}},\ }\bibfield
  {title} {\bibinfo {title} {Variational quantum singular value
  decomposition},\ }\href@noop {} {\bibfield  {journal} {\bibinfo  {journal}
  {Quantum}\ }\textbf {\bibinfo {volume} {5}},\ \bibinfo {pages} {483}
  (\bibinfo {year} {2021}{\natexlab{a}})}\BibitemShut {NoStop}%
\bibitem [{\citenamefont {McClean}\ \emph {et~al.}(2018)\citenamefont
  {McClean}, \citenamefont {Boixo}, \citenamefont {Smelyanskiy}, \citenamefont
  {Babbush},\ and\ \citenamefont {Neven}}]{mcclean2018barren}%
  \BibitemOpen
  \bibfield  {author} {\bibinfo {author} {\bibfnamefont {J.~R.}\ \bibnamefont
  {McClean}}, \bibinfo {author} {\bibfnamefont {S.}~\bibnamefont {Boixo}},
  \bibinfo {author} {\bibfnamefont {V.~N.}\ \bibnamefont {Smelyanskiy}},
  \bibinfo {author} {\bibfnamefont {R.}~\bibnamefont {Babbush}},\ and\ \bibinfo
  {author} {\bibfnamefont {H.}~\bibnamefont {Neven}},\ }\bibfield  {title}
  {\bibinfo {title} {Barren plateaus in quantum neural network training
  landscapes},\ }\href@noop {} {\bibfield  {journal} {\bibinfo  {journal}
  {Nature communications}\ }\textbf {\bibinfo {volume} {9}},\ \bibinfo {pages}
  {4812} (\bibinfo {year} {2018})}\BibitemShut {NoStop}%
\bibitem [{\citenamefont {Cerezo}\ \emph
  {et~al.}(2021{\natexlab{b}})\citenamefont {Cerezo}, \citenamefont {Sone},
  \citenamefont {Volkoff}, \citenamefont {Cincio},\ and\ \citenamefont
  {Coles}}]{cerezo2021cost}%
  \BibitemOpen
  \bibfield  {author} {\bibinfo {author} {\bibfnamefont {M.}~\bibnamefont
  {Cerezo}}, \bibinfo {author} {\bibfnamefont {A.}~\bibnamefont {Sone}},
  \bibinfo {author} {\bibfnamefont {T.}~\bibnamefont {Volkoff}}, \bibinfo
  {author} {\bibfnamefont {L.}~\bibnamefont {Cincio}},\ and\ \bibinfo {author}
  {\bibfnamefont {P.~J.}\ \bibnamefont {Coles}},\ }\bibfield  {title} {\bibinfo
  {title} {Cost function dependent barren plateaus in shallow parametrized
  quantum circuits},\ }\href@noop {} {\bibfield  {journal} {\bibinfo  {journal}
  {Nature communications}\ }\textbf {\bibinfo {volume} {12}},\ \bibinfo {pages}
  {1791} (\bibinfo {year} {2021}{\natexlab{b}})}\BibitemShut {NoStop}%
\bibitem [{\citenamefont {Wang}\ \emph
  {et~al.}(2021{\natexlab{b}})\citenamefont {Wang}, \citenamefont {Fontana},
  \citenamefont {Cerezo}, \citenamefont {Sharma}, \citenamefont {Sone},
  \citenamefont {Cincio},\ and\ \citenamefont {Coles}}]{wang2021noise}%
  \BibitemOpen
  \bibfield  {author} {\bibinfo {author} {\bibfnamefont {S.}~\bibnamefont
  {Wang}}, \bibinfo {author} {\bibfnamefont {E.}~\bibnamefont {Fontana}},
  \bibinfo {author} {\bibfnamefont {M.}~\bibnamefont {Cerezo}}, \bibinfo
  {author} {\bibfnamefont {K.}~\bibnamefont {Sharma}}, \bibinfo {author}
  {\bibfnamefont {A.}~\bibnamefont {Sone}}, \bibinfo {author} {\bibfnamefont
  {L.}~\bibnamefont {Cincio}},\ and\ \bibinfo {author} {\bibfnamefont {P.~J.}\
  \bibnamefont {Coles}},\ }\bibfield  {title} {\bibinfo {title} {Noise-induced
  barren plateaus in variational quantum algorithms},\ }\href@noop {}
  {\bibfield  {journal} {\bibinfo  {journal} {Nature communications}\ }\textbf
  {\bibinfo {volume} {12}},\ \bibinfo {pages} {6961} (\bibinfo {year}
  {2021}{\natexlab{b}})}\BibitemShut {NoStop}%
\end{thebibliography}%

\end{document}